\documentclass[11pt,a4paper]{article}
\usepackage[dvipdfmx]{graphicx}
\usepackage{amsmath, amssymb}

\newcommand{\vct}[1]{\boldsymbol{#1}}
\newcommand{\eden}{\mbox{\Large $\varepsilon$}}
\newcommand{\eb}{E_{\rm B}}

\begin{document}

\baselineskip=16pt

{
\noindent
\Large\textbf{%
The mean-square radius of the neutron distribution in the\\[4pt]
relativistic and non-relativistic mean-field models
}
}

\vspace{\baselineskip}

\noindent
Haruki Kurasawa$^1$\footnote[1]{kurasawa@faculty.chiba-u.jp} and
Toshio Suzuki$^2$\footnote[2]{kt.suzuki2th@gmail.com}

\vspace{0.5\baselineskip}
 
\noindent
$^1$\ \parbox[t]{14cm}{
Department of Physics, Graduate School of Science, Chiba University, \\[2pt]
Chiba 263-8522, Japan
}

\vspace{0.5\baselineskip}

\noindent
$^2$\ \parbox[t]{14cm}{
Research Center for Electron Photon Science, Tohoku University,\\[2pt]
Sendai 982-0826, Japan
}

\vspace{0.5\baselineskip}

\begin{center}
\parbox[t]{14cm}{
\small

\baselineskip=12pt

It is investigated why the root-mean- square radius of the point
neutron distribution is smaller by about 0.1 fm in non-relativistic
mean-field models than in relativistic ones. The difference is shown
to stem from the different values of the product of the effective mass
and the strength of the one-body potential in the two frameworks. 
The values of those quantities are constrained by the Hugenholtz-Van Hove
theorem. The neutron skin is not a simple function of the symmetry
potential, but depends on the nucleon effective mass. 
}
\end{center}

\vspace{0.5\baselineskip}

\section{Introduction}\label{intro}

Recently much has been written on the neutron distribution in nuclei\cite{bm,roca,thi,kss,adh}.
It is one of the most fundamental problems in nuclear physics together with the proton
distribution\cite{bm,deforest}.
The neutron distribution, however, has not been well determined experimentally so far. 
This fact is because the neutron density has been 
studied through hadron probes, where the ambiguity as to the interaction and the reaction 
mechanism is not avoidable yet\cite{thi}.

In contrast to the neutron distribution,
the proton distribution is widely investigated throughout the periodic table of the stable nuclei
theoretically\cite{deforest} and experimentally\cite{vries}. 
The relationship between the point proton and charge density distributions 
is defined unambiguously\cite{deforest,ksmsr}.
The latter is deduced from electron scattering cross sections rather
model-independently\cite{vries}, compared with the strong interaction,
since the electromagnetic interaction is well understood, 
and is so weak that the density distribution of the nuclear ground state is not 
disturbed\cite{deforest, bd}.

It has been believed 
for a long time that electron scattering is useless in the study of the neutron 
distribution in nuclei\cite{deforest,suda}.
Recently, the present authors have proposed a new way to deduce the neutron distribution
from electron scattering data\cite{ksmsr}.
They have derived the exact expression of the $n$th-order moment
of the nuclear charge distribution, and shown that the mean-square radius(msr)
of the charge distribution($R^2_c$) is dominated by the msr of the point proton
distribution($R^2_p$) and is independent of the neutron's msr($R^2_n$),
but that the $n$th-order\,($n\geq 4$) moment of the charge density depends 
on the ($n\!-\!2$)th-order moment of the neutron distribution\cite{ksmsr}.
For example, the fourth-order moment of the charge density($Q^4_c$) depends on $R^2_n$.
Their relationship is uniquely defined, and the value of $Q^4_c$ is well determined
in electron scattering experiment\cite{vries,emrich}. The value of $R^2_n$, however, is not separated
from $Q^4_c$ experimentally.
In order to deduce the value of $R^2_n$ from the experimental value of $Q^4_c$,
it is necessary to rely on a model-dependent analysis.
An advantage to use $Q^4_c$ for deducing the value of $R^2_n$
 is that we need not take care of 
 assumptions on the interaction and reaction mechanism in electron scattering,
 but are able to focus discussions of the model-dependence on nuclear structure.

At present, the nuclear structure is not investigated  without invoking 
phenomenological models. Moreover, most of the
models are constructed for different purposes independently.
Hence, it is not appropriate for the separation of $R^2_n$ from
$Q^4_c$ to choose one model among many existing models.
The present authors\cite{kss} have proposed the least squares analysis(LSA) for the separation
by employing as many previous models as possible together.
Through the LSA, they explore the constraints which are inherent in the framework 
of the nuclear models.
The procedure of the separation is as follows. First $R^2_n$ and $Q^4_c$ are calculated
using several models in the same framework, and then the least-square line(LSL) for those
values is obtained in the $R^2_n-Q^4_c$ plane. Next, the value of $R^2_n$ in the framework
is determined by the cross point of the LSL and the line of $Q^4_c$ corresponding to its 
experimental value.  In order to confirm the obtained result, the LSA of $R^2_n$ against the
other moments also has been performed.
The estimated values of $R^2_n$ are not model-independent, 
but are derived on the basis of the data from the well-known electromagnetic probe,
with utilizing the knowledge on the phenomenological models accumulated for a long time
in nuclear physics. 
A similar method has been proposed for analyzing the parity violating electron scattering\cite{roca},
and employed actually in the analysis of the recent JLab experiment\cite{adh,abra}.

In the Ref.\cite{kss}, the values of $R^2_n$ in $^{40}$Ca, $^{48}$Ca and $^{208}$Pb have been 
estimated, for which the experimental values of $Q^4_c$ are available at present.
They have arbitrary chosen 11 relativistic and 9 non-relativistic 
models among more than 100 versions accumulated for the last 50 years\cite{stone,sw,sw1,fsu}. 
Those models well reproduce fundamental nuclear properties within the mean-field framework,
assuming some nuclei to be a doubly closed shell nucleus.
The LSL is obtained with a small standard deviation and the values of $R^2_n$ are 
determined within the 1$\%$ error including experimental one\cite{kss}.
In this analysis, it has been shown that the relativistic and non-relativistic 
frameworks yield different values of $R^2_n$ from each other in $^{48}$Ca and $^{208}$Pb.
The value of $R_n$ in the non-relativistic models is smaller by about 0.1 fm than that in 
the relativistic models  in both nuclei.
Since in those mean-field models, the values of $R_p$ are fixed so as to reproduce the experimental values,
the neutron skin defined by $\delta R=R_n-R_p$ differs by about 0.1 fm in the two frameworks.
The difference is not between each models, but between the two frameworks,
so that the result is apparently understood to reflect an essential difference
between the structures of the two mean-field approximations.

It should be noted that the 0.1 fm is not small for the neutron skin itself. As seen later, for example,
in $^{208}$Pb, $\delta R$ is 0.275 and 0.162 fm in the relativistic and non-relativistic models,
respectively. Understanding the 0.1 fm difference may be important for the study of 
nuclear fission and fusion phenomena which are sensitive to the structure of nuclear surface\cite{bm,sly}.
Recent detailed calculations\cite{hargen} may not neglect order of 0.1 fm difference in describing
asymmetric nuclei.
The 0.1 fm difference has been pointed out to be crucial for the neutron star
physics also\cite{thi, hargen}.

The purpose of the present paper is to investigate why the values of $R^2_n$ in the non-relativistic
mean-field models is smaller than in the relativistic ones.
The difference will be shown to stem mainly from the difference between the products of the effective mass
and the strength of the one-body potential in the two frameworks.
These two quantities are constrained in each framework by Hugenholtz-Van Hove(HVH)
theorem\cite{bethe, weisskopf, hvh}. The theorem has been proved in the mean-field approximation
in the non-relativistic framework for the symmetric nuclear matter.
As will be shown in the present paper,
the theorem also holds in the asymmetric nuclear matter in both relativistic and non-relativistic models,
and is numerically maintained in the mean-field approximation for finite nuclei also.

In \S\ref{hop}, the root msr $R$ in the one-body potential will be discussed
in order to derive an analytical
expression of $R$ in terms of the strength of the potential and the nucleon effective mass,
using the Woods-Saxon and harmonic potentials.
In \S\ref{relation}, the equations of motion in the relativistic models will be shown to have the
same structure as the Schr\"{o}dinger equation in the non-relativistic models.
In \S\ref{vh},  the HVH theorem will be extended to asymmetric nuclear matter.
In \S\ref{simp}, the complexity of the mean-field models due to a large variety of
interaction parameters will be simplified by using the Woods-Saxon type function,
aiming to make clear the difference between the relativistic and non-relativistic model. 
In \S\ref{hvhlines}, the difference between $\delta R$'s in the two frameworks will be
investigated in detail, according to the HVH theorem.
The final section will be devoted to a brief summary of the present paper.

\section{The nuclear radius in the one-body potential}\label{hop}

Many phenomenological models have been proposed with various interaction parameters\cite{stone,fsu}.
Whether the nuclear radius($R$) is $R_n$ or $R_p$, it may be a complicated function
of their parameters, and the function would be different from one model to another.
The radius, however, is one of the most fundamental quantities which determine the structure of the nucleus,
and, hence, $R_c$ is used as an input to fix the free parameters of the models.
This fact implies that the relationship of $R$ to other key quantities of nuclei
like those in the one-body potential must be almost the same in the mean-field models,
although those key quantities may also depend on the parameters in complicated ways. 

Such relationships of $R$ to other key quantities should hold even in simplified one-body potential models,
if they describe well the gross properties of nuclei\cite{bm}.
As a simple one, Woods-Saxon(WS) potential is most widely used in the literature\cite{bm}.
It may become a guide for the present purpose also, if we have an analytical formula for relationship between
$R$ and the parameters of the one-body Hamiltonian with WS potential,
\begin{equation}
H=\frac{\vct{p}^2}{2M}+V_{\rm ws}(r), \quad
 V_{\rm ws}(r)=\frac{V_{\rm ws}}{1+e^{(r-R_{\rm ws})/a_{\rm ws}}}.\label{h}
\end{equation}
Aiming to have an analytical expression of the relationship, we require the help of the
harmonic oscillator(HO) potential,
\begin{equation}
V_{\rm H}(r)=\frac{k}{2}\left(r^2-R^2_{\rm H}\right), \qquad k=M\omega^2,\label{ho}
\end{equation}
$R_{\rm H}$ being a constant which determines the value of $V_{\rm H}(0)$.
Bohr and Mottelson have shown that the single-particle wave functions in WS potential,
which determine the value of $R$, are well reproduced by those of HO potential\cite{bm}.
In HO potential, the dimension analysis yields
the expression of the radius $R_{\rm ho}$  as
\begin{equation}
 R_{\rm ho}=\frac{C}{(Mk)^{1/4}}\label{hr}
\end{equation}
with $C$ denoting a constant. For the above exact formula, let us search for
the expression of $R_{\rm ho}$ in terms of the WS parameters by minimizing  
the following quantity with respect to the variables, $k$ and $R_{\rm H}$,
\begin{equation}
F_n(k, R_{\rm H})= \int^{R_{\rm H}}_0dr\,r^n\Bigl( V_{\rm H}(r)
			  -V_{\rm ws}(r) \Bigr)^2, \quad\label{wmin}
 \end{equation}
where $n$ is considered to be 0 for the surface integral and 2 for the volume integral.
The value of $n$ is chosen by referring to Ref.\cite{bm},
which shows a similarity of the wave functions in the two
potentials with $\omega=8.6$ MeV and $V_{\rm H}(0)=-55$ MeV and with
 $V_{\rm ws}=-50$ MeV, $R_{\rm ws}=5.8$ fm and $a_{\rm ws}=0.65$ fm.
The numerical method yields the minimum values of $F_n$ for the same WS parameters at
$\omega=8.63$ MeV and $V_{\rm H}(0)=-55.20$ MeV for $n=0$, and at $\omega=9.66$ MeV
and $V_{\rm H}(0)=-62.65$ MeV for $n=2$. Comparing these values with those in Ref.\cite{bm},
it may be reasonable to employ $n=0$, rather than $n=2$, for reproducing the wave functions
in WS potential.

Once we determine the value of $n$, it is possible to derive the analytical formula for the
approximate relationship between $R_{\rm ho}$ and the WS parameters.
Eq.(\ref{wmin}) for $n=0$ is written as
\begin{equation}
F_0(k, R_{\rm H})= \int^{R_{\rm H}}_0dr\Bigl( V_{\rm H}(r) -V_{\rm ws}(r) \Bigr)^2
 =\frac{2}{15}k^2R_{\rm H}^5+F_{\rm H}+F_{\rm V},\nonumber
 \end{equation}
where we have defined  
\begin{align}
F_{\rm H}&=-2\int_0^{R_{\rm H}}dr\,V_{\rm H}(r)V_{\rm ws}(r)
 =-kV_{\rm ws}\int_0^\infty dr\,\frac{r^2-R^2_{\rm H}}{1+e^{(r-R_{\rm ws})/a_{\rm ws}}}+\delta F_{\rm H},
 \label{fh}\\[2pt]
F_{\rm V}&=\int_0^{R_{\rm H}}dr\,V^2_{\rm ws}(r)
 =V^2_{\rm ws}\int_0^\infty \frac{dr}{\left(1+e^{(r-R_{\rm ws})/a_{\rm ws}}\right)^2}+\delta F_{\rm V},
 \label{fv} 
\end{align}
with
\begin{equation}
\delta F_{\rm H}=kV_{\rm ws}\int_{R_{\rm H}}^\infty dr\,
 \frac{r^2-R^2_{\rm H}}{1+e^{(r-R_{\rm ws})/a_{\rm ws}}},\qquad
\delta F_{\rm V}=-V^2_{\rm ws}\int_{R_{\rm H}}^\infty 
 \frac{dr}{\left(1+e^{(r-R_{\rm ws})/a_{\rm ws}}\right)^2}.
\end{equation}
Using the identity for a general function $g(r)$,
\[
 \int_{R_{\rm H}}^\infty dr\,
 \frac{g(r)}{\left(1+e^{(r-R_{\rm ws})/a_{\rm ws}}\right)^n}
 =\varDelta^na_{\rm ws}\int_0^\infty dx\, \frac{g(a_{\rm ws}x+R_{\rm H})e^{-nx}}{(1+\varDelta e^{-x})^n},
 \quad \varDelta= e^{-(R_{\rm H}-R_{\rm ws})/a_{\rm ws}},
\]
we can neglect $\delta F_{\rm H}$ and $\delta F_{\rm V}$ in Eq.(\ref{fh}) and (\ref{fv}),
assuming $\varDelta \ll 1$.
Then, $F_{\rm V}$ in Eq.(\ref{fv}) is independent of $k$ and $R_{\rm H}$,
and it is enough to minimize the only first term of the most right-hand side of Eq.(\ref{fh}).
The integral of the first term is performed with the use of Sommerfeld expansion.
In neglecting contributions of relative order $e^{-R_{\rm ws}/a_{\rm ws}}$\cite{bm},
it is written as
\[
 \int_0^\infty dr\,\frac{g(r)}{1+e^{(r-R_{\rm ws})/a_{\rm ws}}}=\int_0^{R_{\rm ws}}dr\,g(r)
 +\frac{\pi^2a^2_{\rm ws}}{6}g'(R_{\rm ws})+\frac{7\pi^4a^4_{\rm ws}}{360}g'''(R_{\rm ws}) +\cdots.
\]
Since $g'''(r)=0$ for Eq.(\ref{fh}), we have 
\[
F_0(k, R_{\rm H})=\frac{2}{15}k^2R^5_{\rm H}+kV_{\rm ws}\left(R^2_{\rm H}R_{\rm ws}
 -\frac{1+b_{\rm ws}}{3}R^3_{\rm ws}\right),
\quad b_{\rm ws}=\left(\frac{\pi a_{\rm ws}}{R_{\rm ws}}\right)^2.
\]
It should be noticed that there is no higher-order contribution from the diffuseness parameter.
The partial differentials of the above equation $F_0$ with respect to $k$ and $R_{\rm H}$ yield
its minimum value at 
\begin{align}
k&=-3\left(\frac{3}{5}\right)^{3/2}\frac{V_{\rm ws}}{R^2_{\rm ws}(1+b_{\rm ws})^{3/2}},\label{hk}\\[2pt]
R^2_{\rm H}
&=\frac{5}{3}(1+b_{\rm ws})R^2_{\rm ws}, \qquad 
V_{\rm H}(0)=\frac{3}{2}\sqrt{\frac{3}{5}}\frac{V_{\rm ws}}{\sqrt{1+b_{\rm ws}}}.\nonumber
\end{align}
When employing the values, $V_{\rm ws}=-50$ MeV, $R_{\rm ws}=5.8$ fm and $a_{\rm ws}=0.65$ fm
in Ref.\cite{bm}, the above equations provide $\omega=8.49$ MeV and $V_{\rm H}(0)=-54.80$ MeV,
which reproduce almost the same values obtained by the numerical method mentioned above.

Finally, inserting Eq.(\ref{hk}) into Eq.(\ref{hr}), $R_{\rm ho}$
is described approximately in terms of the WS parameters as 
\begin{equation}
R_{\rm ho}\approx B\left(-\frac{R^2_{\rm ws}}{m^\ast V_{\rm ws}}\right)^{1/4}
 \left(1+b_{\rm ws}\right)^{3/8},\label{ra}
\end{equation}
$B$ being a constant. In the above equation, the nucleon mass has been replaced by the effective mass,
$M^\ast=Mm^\ast$.
Eq.(\ref{ra}) expresses well our expectation such that the value of $R$ increases with $R_{\rm ws}$,
and decreases with increasing $(-V_{\rm ws})$ and $m^\ast$.
Indeed, the first parenthesis of the right-hand side
may be derived in the square-well potential with the depth $V_{\rm ws}$ and the width $R_{\rm ws}$.
Eq(9) shows that the diffuseness parameter contributes to the radius in the form of
$(a_{\rm ws}/R_{\rm ws})^2$.

If the neutron potential, $V_n$, and effective mass, $m_n^\ast$, are different from
$V_p$ and $m^\ast_p$ of the proton, the value of $R_n$ may be different from that of $R_p$.
In the same way, if $V_n$ and $m_n^\ast$ in the one model are different
from those in another model,  their $R_n$'s are different from each other.
When comparing the nuclear radius, $R_1$ in the one framework with $R_2$ in another one,
the following expression is useful,
\begin{equation}
\frac{R_1}{R_2}=\left(\frac{m_2^\ast V_{\rm ws,2}}{m^\ast_1 V_{\rm ws,1}}\right)^{1/4}
 \left(\frac{R_{\rm ws,1}}{R_{\rm ws,2}}\right)^{1/2}
 \left(\frac{1+b_{\rm ws,1}}{1+b_{\rm ws,2}}\right)^{3/8}.\label{rra}
\end{equation}

\section{Equations of motion of the mean-field models}\label{relation}

Eq.(\ref{ra}) and (\ref{rra}) are simple enough to understand the relationship
between $R_{\rm ho}$ and the key quantities of the one-body potential.
The effective mass and the one-body potential are well-defined quantities
in the mean-field models.
Expecting that such a simple relationship holds approximately in those
phenomenological models also, let us investigate how they appear in the equations of motion
in the relativistic and non-relativistic models.
 
In the relativistic nonlinear $\sigma-\omega-\rho$ model, the nuclear Lagrangian is given,
using the notations in the literature\cite{sw,fsu,nl3}, by
\begin{align}
\mathcal{L}&=\overline{\psi}\left(i\gamma_\mu\partial^\mu-M-g_\sigma\sigma-g_\omega\gamma_\mu\omega^\mu
 -g_\rho\gamma_\mu\vct{\tau}\!\cdot\!\vct{b}^\mu-e\gamma_\mu A^\mu\frac{1+\tau_3}{2}\right)\psi \nonumber\\
&\hphantom{=}+\frac{1}{2}(\partial_\mu\sigma)^2-\frac{m^2_\sigma}{2}\sigma^2-\frac{g_3}{3}\sigma^3
-\frac{g_4}{4}\sigma^4 -\frac{1}{4}\omega_{\mu\nu}\omega^{\mu\nu}+\frac{m^2_\omega}{2}\omega_\mu\omega^\mu
 +\frac{c_4}{4}(\omega_\mu\omega^\mu)^2\nonumber\\
&\hphantom{=}-\frac{1}{4}\vct{b}_{\mu\nu}\!\cdot\!\vct{b}^{\mu\nu}
+\frac{m^2_\rho}{2}\vct{b}_\mu\!\cdot\!\vct{b}^\mu
+\lambda g^2_\rho\vct{b}_\mu\!\cdot\!\vct{b}^\mu g^2_\omega\omega_\nu\omega^\nu
-\frac{1}{4}A_{\mu\nu}A^{\mu\nu}.
\end{align}
Then, Euler-Lagrange equation provides us with the equations of motion for the static mean-field, 
\begin{align}
&\Bigl(-i\vct{\alpha}\cdot\vct{\nabla}+\gamma_0(M+V_\sigma)+V_0\Bigr)\psi=(E+M)\psi, \label{eqm}\\[2pt]
&\left(-\vct{\nabla}^2+m^2_\sigma\right)V_\sigma=-g^2_\sigma\left(\rho_S+\frac{g_3}{g^3_\sigma}V^2_\sigma
 +\frac{g_4}{g^4_\sigma}V^3_\sigma\right), \\[2pt]
&\left(-\vct{\nabla}^2+m^2_\omega \right)V_\omega=g^2_\omega\left(\rho-\frac{c_4}{g^4_\omega}V^3_\omega
 -2\lambda V_\omega V^2_\rho\right), \\[2pt]
&\left(-\vct{\nabla}^2+m^2_\rho \right)V_\rho
=g^2_\rho\bigl(\rho_p-\rho_n-2\lambda V^2_\omega V_\rho\bigr),\\[2pt]
&-\vct{\nabla}^2V_c=e^2\rho_p.\label{rho}
\end{align}
In the above equations from Eq.(\ref{eqm}) to (\ref{rho}), we have defined $\psi$ as
a single-particle wave function, and used following notations,
$V_\sigma=g_\sigma\sigma, V_\omega=g_\omega\omega^0, V_\rho=g_\rho b_3^0$
and $V_c$ for the Coulomb potential, $V_c=eA^0$.
Moreover, $V_0$ is given by
\begin{equation}
  V_0(\vct{r})=V_\omega(\vct{r})+V_\rho(\vct{r})\tau_3 + V_c(\vct{r})
 \frac{1+\tau_3}{2}
\end{equation}
with $\tau_3=+1(-1)$ for protons(neutrons), and the nucleon densities are
\[
 \rho_{S}(\vct{r})=\sum_{\alpha}
\overline{\psi}_\alpha(\vct{r})\psi_\alpha(\vct{r}),\quad
\rho_\tau(\vct{r})=\sum_{\alpha\in \tau}
\psi_\alpha^\dagger(\vct{r})\psi_\alpha(\vct{r}),\quad \rho(\vct{r})=\rho_n(\vct{r})+\rho_p(\vct{r}),
\]
with $\tau=p$ for protons and $\tau=n$ for neutrons.

Eq.(\ref{eqm}) represents the two coupled equations for the upper component, $\psi_u(\vct{r})$,
and the lower two component, $\psi_d(\vct{r})$, of $\psi(\vct{r})$. One of them gives 
\begin{equation}
\psi_d(\vct{r})
 =-\,\frac{1}{2M^\ast_\tau(\vct{r})}i\vct{\sigma}\!\cdot\!\vct{\nabla}\psi_u(\vct{r})\,,
 \label{ud}
\end{equation}
writing the effective nucleon mass, $ M^\ast_{\tau}(\vct{r})$, as
\begin{equation}
  M^\ast_{\tau}(\vct{r})=\frac{2M+E+V_\sigma(\vct{r})-V_0(\vct{r})}{2}.\label{effective}
\end{equation}
In inserting Eq.(\ref{ud}) into the other equation
of Eq.(\ref{eqm}), we obtain the Schr\"{o}dinger-like equation as
\begin{equation}
\left(-\vct{\nabla}\frac{1}{2M^\ast_\tau(\vct{r})}\!\cdot\!\vct{\nabla}+V_\tau(\vct{r})
+V_c(\vct{r})\frac{1+\tau_3}{2}-i\left(\vct{\nabla}\frac{1}{2M^\ast_\tau(\vct{r})}
\right)\!\cdot\!(\vct{\nabla}\!\times\!\vct{\sigma})\right)\psi_u(\vct{r})=E\psi_u(\vct{r}).
\label{rse}
\end{equation}
In the above equation, the nuclear potential, $V_\tau(\vct{r})$, is defined by
\begin{equation}
V_\tau(\vct{r})=V_\sigma(\vct{r})+V_\omega(\vct{r})+V_\rho(\vct{r})\tau_3.\label{rv}
\end{equation}
We note that the effective mass, $M^\ast_\tau(\vct{r})$, is written approximately as
\begin{equation}
M^\ast_{\tau}(\vct{r})\approx M+\frac{1}{2}\left(V_\sigma(\vct{r})-V_\omega(\vct{r})
   -V_\rho(\vct{r})\tau_3-V_c(\vct{r})\frac{1+\tau_3}{2}\right),
   \label{reff}
\end{equation}
using the fact that $2M+E\approx 2M$.
For $^{208}$Pb, the values of the potentials around the center of the nuclear density
are about $V_\sigma \approx -380$ MeV, $V_\omega \approx 306$ MeV and $V_\rho \approx -6$
MeV\cite{nl3}. It should be noted that the effective mass in the relativistic models is almost isoscalar,
and is dominated by $V_\sigma$ and $V_\omega$ in the same way as the spin-orbit potential
in the last term of the left-hand side in Eq.(\ref{rse}).

The root-msr's of the point proton and neutron distributions calculated with NL3\cite{nl3}
are listed in Table 1. They are defined as
\begin{align*}
R^2_\tau
&=\frac{1}{N_\tau}\sum_{\alpha \in \tau}\int^\infty_0 dr\,r^2
 \Bigl( G^2_\alpha(r)+F^2_\alpha(r) \Bigr)\,, \qquad 
(R_\tau)^2_G=\frac{1}{N_\tau}\sum_{\alpha \in \tau}\int^\infty_0 dr\, r^2
 G^2_\alpha(r)\,, \\[2pt]
(R_\tau)^2_N
&=\frac{1}{N_\tau}\sum_{\alpha \in \tau}\int^\infty_0 dr\,r^2
 \frac{G^2_\alpha(r)}{n_G}\,,\qquad n_G=\int^\infty_0 dr\, G^2_\alpha(r)\,,
\end{align*}
where $G_\alpha(r)/r$ and $F_\alpha(r)/r$ denote the radial part of the large and the small
component of $\psi_\alpha(\vct{r})$, respectively, with the normalization,
$\int^\infty_0dr\bigl(G^2_\alpha(r)+F^2_\alpha(r)\bigr)=1$.
Moreover, we have defined $N_\tau=N(Z)$ for $\tau=n(p)$,
and $n_G$ for the normalization of the upper component used in $(R_\tau)^2_N$.
Table 1 also shows the ratios of  $(R_\tau)_G$ and  $(R_\tau)_N$ to $R_\tau$ in the
parentheses. As seen from $(R_\tau)_G$ in Table 1, the contribution of the lower component to $R_\tau$
is about $1\%$, and it is absorbed into $(R_\tau)_N$ which is calculated with the renormalized
large component $G_\alpha(r)/\sqrt{n_G}$. Similar results are obtained in other relativistic models.
According to these results, we will use the renormalized large component,
ignoring the small component, when comparing the relativistic models
with the non-relativistic ones below.

We note that in principle, the two-component framework equivalent to the four-component one should be
derived by the Foldy-Wouthuysen unitary transformation\cite{bd}.
In order to obtain the normalized two component wave functions,
Eq.(\ref{rse}) will be used only in the present paper for comparison with non-relativistic models,
for simplicity and transparency.
In Ref.\cite{kss}, the calculations of the msr in the relativistic models have been performed
within the four-component framework.

\begin{table}
\begingroup
\renewcommand{\arraystretch}{1.2}
{\setlength{\tabcolsep}{4pt}
\hspace{0.5cm}\begin{tabular}{|c|c|c|c||c|c|c|} \hline
 & $R_n$ & $(R_n)_G$ & $(R_n)_N$ & 
   $R_p$ & $(R_p)_G$ & $(R_p)_N$ \\ \hline 
$^{48}$Ca  & $3.6050$ & $3.5736(0.991)$ & $3.6082(1.001)$  & $3.3789$ & $3.3522(0.992)$ & $3.3846(1.002)$  \\ 
$^{208}$Pb & $5.7405$ & $5.6888(0.991)$ & $5.7522(1.002)$  & $5.4600$ & $5.4135(0.991)$ & $5.4656(1.001)$  \\ \cline{1-7}
\end{tabular}
}
\endgroup
\caption{ The root msr of the point neutron($R_n$) and proton($R_p$) distribution calculated with NL3 for
$^{48}$Ca and $^{208}$Pb. The number is given in units of fm, except for the one in the parenthesis
which denotes the ratio to the $R_\tau$. For details, see the text.}
\label{table_1}
\end{table}

In the Skyrme Hartree-Fock approximation in the non-relativistic models,
the Schr\"{o}dinger equation is written as \cite{gian,sly4},
\begin{equation}
\left(-\vct{\nabla}\frac{1}{2M^\ast_\tau(\vct{r})}\!\cdot\!\vct{\nabla}+V_\tau(\vct{r})
+V_c(\vct{r})\frac{1+\tau_3}{2}-i\vct{W}_\tau(\vct{r})\!\cdot(\vct{\nabla}\!\times\!\vct{\sigma})
\right)\varphi(\vct{r})=E\varphi(\vct{r}),
\label{nse}
\end{equation}
where using the same notations as in Ref.\cite{sly4}, $M^\ast_\tau(\vct{r})$, $V_\tau(\vct{r})$
and $\vct{W}_\tau(\vct{r})$ are given as,
{\allowdisplaybreaks
\begin{align}
\frac{1}{M^\ast_\tau(\vct{r})}
&=\frac{1}{M}+\frac{t_1(2+x_1)+t_2(2+x_2)}{4}\rho(\vct{r})
+\frac{t_2(1+2x_2)-t_1(1+2x_1)}{4}\rho_\tau(\vct{r})\,,\label{neff}\\[4pt]
%%%%%
V_\tau(\vct{r})
&=\frac{t_0}{2}\Bigl( (2+x_0)\rho(\vct{r})-(1+2x_0)\rho_\tau(\vct{r}) \Bigr)
+\frac{t_3}{24} (2+x_3)(2+\alpha)\rho^{\alpha+1}(\vct{r}) \nonumber \\[2pt]
&
\hphantom{=}
-\frac{t_3}{24} 
(2x_3+1)\Bigl[2\rho^\alpha(\vct{r})
 \rho_\tau(\vct{r})+\alpha\rho^{\alpha-1}(\vct{r})
\Bigl( \rho^2_p(\vct{r}) +\rho^2_n(\vct{r}) \Bigr) \Bigr] \nonumber\\[2pt]
&
\hphantom{=}
+\frac{t_1(2+x_1)+t_2(2+x_2)}{8}K(\vct{r})+\frac{t_2(1+2x_2)-t_1(1+2x_1)}{8}K_\tau(\vct{r}) \nonumber\\[2pt]
&
\hphantom{=}
+\frac{t_2(2+x_2)-3t_1(2+x_1)}{16}\vct{\nabla}^2\rho(\vct{r})+\frac{3t_1(1+2x_1)+t_2(1+2x_2)}{16}
 \vct{\nabla}^2\rho_\tau(\vct{r})\nonumber\\
&
\hphantom{=}
-\frac{W_0}{2}\vct{\nabla}\!\cdot\!\bigl(\vct{J}(\vct{r})+\vct{J}_\tau(\vct{r})\bigr)\,,\label{nv}\\[4pt]
%%%%%%
\vct{W}_\tau(\vct{r})
&=\frac{W_0}{2}\vct{\nabla}\bigl( \rho(\vct{r})+\rho_\tau(\vct{r}) \bigr)
 +\frac{t_1-t_2}{8}\vct{J}_\tau(\vct{r})-\frac{t_1x_1+t_2x_2}{8}\vct{J}(\vct{r}) \nonumber \\[2pt]
&\approx \frac{W_0}{2}\vct{\nabla}\bigl( \rho(\vct{r})+\rho_\tau(\vct{r}) \bigr).\label{nls}  
\end{align}}
In Eq.(\ref{nv}), $K(\vct{r})=K_n(\vct{r})+K_p(\vct{r})$ has been defined
with $K_\tau(\vct{r})=\sum_{\alpha\in\tau}
|\vct{\nabla}\varphi_\alpha(\vct{r})|^2$,
and $\vct{J}(\vct{r})=\vct{J}_n(\vct{r})+\vct{J}_p(\vct{r})$, where $\vct{J}_\tau(\vct{r})$ denotes
the spin density given in Ref.\cite{sly4}.

It is seen that Eq.(\ref{nse}) in the non-relativistic models has the same structure
as Eq.(\ref{rse}) in the relativistic models.
They are composed of the four parts, $M^\ast_\tau(\vct{r})$, $V_\tau(\vct{r})$,
$V_c(\vct{r})$ and the spin-orbit potential.
If the strengths and the coordinate-dependences of these parts were the same in the two frameworks,
one could not distinguish one framework from another, in spite of their complicated parameter sets.
Among the four parts, the last two ones are expected to play a minor role
in the present purpose to explore the difference between $\delta R$'s in the two frameworks.
The Coulomb potential is almost the same, and the strengths of the spin-orbit potentials
reproduce experimental values of the single-particle energy levels in both frameworks\cite{sw,sk1}.
In contrast to these, the first two parts are strongly model-dependent.
The values of the effective masses are spread out over a wide range \cite{stone}.
Similarly, there is no reason why the one-body potentials are almost the same in all the mean-field models.
Hence, the 0.1 fm difference between $\delta R$'s may be related
to $M^\ast_\tau(\vct{r})$ and $V_\tau(\vct{r})$ depending on the different interaction parameters.

This observation is consistent with Eq.(\ref{ra}) and (\ref{rra}) which clearly indicate
that the difference problem is related to the effective mass and one-body potential.
It is also apparent that they are not independent of each other.
On the one hand, the product of the $M^\ast_p(\vct{r})$ and $V_p(\vct{r})$ is constrained
by hand so as to reproduce the experimental value of $R_c$ in both relativistic
and non-relativistic models. On the other hand, there is not a similar constraint
on the neutron distribution, but both frameworks predict the values of $R_n$ which are
distributed within a narrow range around each average value\cite{kss}.
If the difference between $\delta R$'s is actually related to the effective mass and the one-body
potential, there should be another constraint on the variations of these two quantities,
which works differently in the relativistic and non-relativistic models.

As one of such candidates, it may be natural to expect the symmetry energy\cite{roca}.
The symmetry energy coefficient, $a_4$\cite{ver}, is composed of the potential 
and kinetic parts\cite{bm}, which are given in the present
relativistic and non-relativistic mean-field models,
respectively, as\cite{stone}
\begin{align}
a_{4,{\rm rel}}&=\frac{k_{\rm F}^2}{6\sqrt{k^2_{\rm F}+M^ 2_\sigma}}+\frac{\rho}{2}\frac{g^2_\rho}
{m^2_\rho+2\lambda g^2_\rho V^2_\omega},\,\qquad  M_\sigma=M+V_\sigma, \label{rs}\\[4pt]
a_{4,{\rm non}}&= \left(\frac{3\pi^2}{2}\right)^{2/3}\left(\frac{\rho^{2/3}}{6M}
 +\frac{-3t_1x_1+t_2(5x_2+4)}{24}\rho^{5/3}\right) \nonumber \\[2pt]
&
\hphantom{=}
 -\frac{2x_0+1}{8}t_0\rho-\frac{2x_3+1}{48}t_3\rho^{\alpha+1},\label{ns} 
\end{align}
where $k_{\rm F}$ denotes the Fermi momentum, and $\rho$ the nucleon density in the nuclear matter.
Actually, they are related to the difference between the neutron and proton potentials
in Eq.(\ref{rv}) and (\ref{nv}), and the effective mass in Eq.(\ref{effective}) and (\ref{neff}).
The relationship of $a_4$ to $\delta R$, however, does not seem to be described explicitly.
In fact, there is more fundamental restriction on the relationship between
the potential and the effective mass. 
It is known as the Hugenholtz-Van Hove(HVH) theorem\cite{bethe,weisskopf,hvh},
which holds in any mean-field model for symmetric nuclear matter.

\section{Hugenholtz-Van Hove theorem}\label{vh}

According to the HVH theorem,
the binding energy per nucleon is equal to the Fermi energy in symmetric nuclear matter.
Both relativistic and non-relativistic models have been constructed so as to satisfy the
theorem at the values of the binding energy of the nucleon to be about $-16$MeV
and of the Fermi momentum to be about 1.3 fm$^{-1}$.
These values are used as inputs in order to fix their free parameters in the nuclear interactions.
The Fermi energy is given by the sum of the kinetic and potential energy, so that
the strength of the potential and the value of the effective mass are constrained by these inputs.
Since the HVH theorem has been proved for the only symmetric nuclear matter\cite{bethe,weisskopf,hvh},
however, we will extend the theorem to the relativistic and non-relativistic asymmetric nuclear matter,
and utilize the theorem as a guide of the analysis of $\delta R$ in neutron-rich finite nuclei.

\subsection{ HVH theorem in the symmetric nuclear matter}\label{snm}

Hugenholtz and Van Hove have shown that the following equation holds
in the non-relativistic mean-field model for symmetric nuclear matter\cite{bethe,weisskopf,hvh},
\begin{equation}
\frac{\eden}{\rho} = E_{\rm F},
 \quad {\rm when} \quad \frac{d}{d\rho}\frac{\eden}{\rho}=0,
 \label{hvh}
\end{equation}
where $\eden$ stands for the total energy density of the system,
and $E_{\rm F}$ the Fermi energy. The value of $\eden/\rho$ represents
the binding energy per nucleon, $\eb$, to be written in the non-relativistic models, as
\begin{equation}
\eb=E_{\rm F}=\frac{k^2_{\rm F}}{2m^\ast M}+V. \label{be}
\end{equation}
In the relativistic models, $\eden/\rho$ and $E_{\rm F}$ contain the nucleon rest mass.
Hence, $\eb$ and $E_{\rm F}$ are given by
\[
\eb=\frac{\eden}{\rho} -M = E_{\rm F}-M\,.
\]
In the present relativistic models, $E_{\rm F}$ in the symmetric nuclear matter is written as\cite{sw}
\[
E_{\rm F}=\sqrt{k^2_{\rm F}+M^2_\sigma} + V_\omega, \qquad M_\sigma=M+V_\sigma .
\]
In setting
\[
 \eb=K+V, \quad K=\sqrt{k^2_{\rm F}+M_\sigma^ 2}-M_\sigma, \quad V=V_\sigma+V_\omega,
\]
$K$ is described as
\begin{equation}
K=\frac{k^2_{\rm F}}{2M^\ast}(1-\delta)\approx \frac{k^2_{\rm F}}{2M^\ast},
\end{equation}
with $M^\ast=M+\bigl( V_\sigma-V_\omega\bigr)/2$ from Eq.(\ref{reff}).
We have defined
\begin{equation}
\delta =1-\frac{2M^\ast}{k^2_{\rm F}}\left(\sqrt{k^2_{\rm F}+M^2_\sigma}-M_\sigma\right),
\label{d}
\end{equation}
and used the fact that $\delta \ll 1$ in taking the values of Ref.\cite{nl3} for the right-hand side.
Thus, in the relativistic models also, $\eb$ is expressed in the form as Eq.(\ref{be}).
Finally, in both relativistic and non-relativistic models, the nuclear potential is
inversely proportional to the effective mass, according to the HVH theorem.
In the case of Eq.(\ref{be}), we have
\begin{equation}
V=\frac{a}{m^\ast} + b, \label{ab}
\end{equation}
where $a\approx -35$ MeV and $b\approx -16$ MeV for $k_{\rm F}\approx 1.3$ fm$^{-1}$ and
$\eb\approx -16$ MeV.

Indeed, it is verified that all the relativistic and non-relativistic models employed
in the present paper satisfy Eq(\ref{hvh}) explicitly. In the non-relativistic models,
we have for the protons and neutrons, separately,
\begin{equation}
 \frac{\partial\eden}{\partial\rho_\tau}=\frac{k^2_{{\rm F}\tau}}{2M^\ast_\tau}
  +V_\tau=E_{{\rm F}\tau},
\label{non}
\end{equation}
while in the relativistic $\sigma-\omega-\rho$ models,
\begin{equation}
 \frac{\partial\eden}{\partial\rho_\tau}=\sqrt{k^2_{{\rm F}\tau}+M_\sigma^ 2}
 +V_\omega+\tau V_\rho=E_{{\rm F}\tau}. \label{rel}
\end{equation}
In the above equations, the total energy density in the non-relativistic models is written as\cite{gian,sly4}
\begin{align*}
 \eden
&=\frac{K_p}{2M^\ast_p}+\frac{K_n}{2M^\ast_n}+\frac{t_0}{4}
 \Bigl( (2+x_0)\rho^2-(2x_0+1)(\rho^2_p+\rho^2_n) \Bigr) \\[2pt]
&
\hphantom{\,=\,\frac{K_p}{2M^\ast_p}+\frac{K_n}{2M^\ast_n}}
+\frac{t_3}{24}\Bigl( (2+x_3)\rho^{\alpha+2}-(2x_3+1)\rho^\alpha(\rho^2_p+\rho^2_n) \Bigr),
\end{align*}
where we have defined $K_\tau=3k^2_{{\rm F}\tau}\rho_\tau/5$ with $k_{{\rm F}\tau}=(3\pi^2\rho_\tau)^{1/3}$.
In the relativistic models, it is given by
\[
 \eden=\eden_K+V_\omega\rho+V_\rho(\rho_p-\rho_n)+U_\sigma-U_0,
\]
using the abbreviations,
\begin{align}
\eden_K&= \frac{2}{(2\pi)^3}\int^{k_{{\rm F}p}}_0d^3k\sqrt{k^2+M^2_\sigma}
 +\frac{2}{(2\pi)^3}\int^{k_{{\rm F}n}}_0d^3k\sqrt{k^2+M^2_\sigma},\nonumber\\[2pt]
U_\sigma&=\frac{m^2_\sigma}{2g^2_\sigma}V^2_\sigma+ \frac{g_3}{3g^3_\sigma}V^3_\sigma
 +\frac{g_4}{4g^4_\sigma}V^4_\sigma,
 \qquad U_0=\frac{m^2_\omega}{2g^2_\omega}V^2_\omega+ \frac{c_4}{4g^4_\omega}V^4_\omega
 +\frac{m^2_\rho}{2g^2_\rho}V^2_\rho
 +\lambda V^2_\omega V^2_\rho,
\end{align}
which satisfy the equations of motion for the mesons,
\begin{align*}
 \frac{\partial U_\sigma}{\partial V_\sigma}
&=\frac{m^2_\sigma}{g^2_\sigma}V_\sigma
 +\frac{g_3}{g^3_\sigma}V^2_\sigma+\frac{g_4}{g^4_\sigma}V^3_\sigma=-\rho_S, \qquad
 \rho_S = \frac{\partial \eden_K}{\partial M_\sigma}\,, \\[2pt]
 \frac{\partial U_0}{\partial V_\omega}
&=\frac{m^2_\omega}{g^2_\omega}V_\omega
 +\frac{c_4}{g^4_\omega}V^3_\omega+2\lambda V_\omega V^2_\rho=\rho\,,\qquad
 \frac{\partial U_0}{\partial V_\rho}=\frac{m^2_\rho}{g^2_\rho}V_\rho
 +2\lambda V^2_\omega V_\rho=\rho_p-\rho_n.
 \end{align*}

In both relativistic and non-relativistic models, 
$\partial\eden/\partial\rho_\tau=E_{{\rm F}\tau}$ holds at any value of $\rho$ or $k_{\rm F}$,
so that we have for $\rho_p=\rho_n=\rho/2$,  
\[
 \frac{d\eden}{d\rho}=\frac{1}{2}\left(\frac{\partial \eden}{\partial \rho_p}
 +\frac{\partial \eden}{\partial \rho_n}\right)=\frac{E_{{\rm F}n}+E_{{\rm F}n}}{2}=E_{\rm F},
 \qquad
 \rho\frac{d}{d\rho}\frac{\eden}{\rho}=\frac{d\eden}{d\rho}-\frac{\eden}{\rho}
 =E_{\rm F}-\frac{\eden}{\rho},
\]
as should be.
The last equation yields Eq.(\ref{hvh}) for $(d/d\rho)(\eden/\rho)=0$.
Thus, in the mean-field models, Eq.(\ref{non}) and (\ref{rel}),
which hold for protons and neutrons separately,
are essential for the HVH theorem to be valid.

\subsection{ HVH theorem in the asymmetric nuclear matter}\label{asnm}

In order to discuss neutron-rich nuclei using the HVH theorem as a guide,
we have to extend the theorem so as to be applicable to the relativistic and non-relativistic
asymmetric nuclear matter.

One of the naive ways to the extension may be to minimize the total energy per nucleon,
assuming $\rho_n=\nu\rho$ and $\rho_p=(1-\nu)\rho$ for a fixed value of $\nu$\cite{bm,stone,sly4}.
This choice is not, however, appropriate for the present purpose,
since $ E_{{\rm F}n}$ and $E_{{\rm F}p}$ remain as in Eq.(\ref{non}) and (\ref{rel})
without the Coulomb energy.
Moreover, if $\rho_n(r)=\nu\rho(r)$
and $\rho_p(r)=(1-\nu)\rho(r)$
were realized in finite nuclei,
one would have $\delta R=0$ even for $N\ne Z$ nuclei.
In order to extend the HVH theorem for asymmetric nuclear matter,
it is be better to avoid these defects.
For this purpose, without using the parameter $\nu$,
we make a model for the neutron and proton system in taking account effects of
the ``Coulomb potential'' explicitly as below.

We require for asymmetric nuclear matter
\begin{equation}
\frac{\partial}{\partial\rho_\tau}\frac{\eden_{\rm asym}}{\rho}=0,\label{me}
\end{equation}
adding $v_c\rho_p$ as the ``Coulomb term'' to the total energy density\cite{brack},
\begin{equation}
\eden_{\rm asym}=\eden+v_c\rho_p, \label{med}
\end{equation}
where $v_c$ is a constant.
The above equation is assumed in order to make a model
which may be used just as a guide for the following discussions on the stable finite nucleus
where Fermi energies of neutrons and protons are the same and the Coulomb
potential is necessary. We will see later that the final results of this paper listed in Table 12
do not depend on the above form of the ``Coulomb term'' and its strength $v_c$.
Then, since Eq.(\ref{non}) and (\ref{rel}) still hold, we have the expression of the binding energy,
\begin{equation}
\eb=E_{{\rm F}\tau}=\frac{k^2_{{\rm F}\tau}}{2M^\ast_\tau}+V_\tau
 +\frac{1+\tau_3}{2}v_c, \label{nonb}
\end{equation}
in the non-relativistic models, while in the relativistic models,
\begin{align}
\eb=E_{{\rm F}\tau}-M
&=\sqrt{k^2_{{\rm F}\tau}+M^2_\sigma}-M_\sigma+V_\tau+\frac{1+\tau_3}{2}v_c \nonumber \\
&=(1-\delta_\tau)\frac{k^2_{{\rm F}\tau}}{2M^\ast_\tau}+V_\tau+\frac{1+\tau_3}{2}v_c, \label{relb}
\end{align}
where $\delta_\tau$ is given by Eq.(\ref{d}) with $k_{{\rm F}\tau}$ and $M^\ast_{\tau}$ instead of $k_{\rm F}$
and $M^\ast$, respectively,
while $V_\tau$ and $M^\ast_\tau$ are given by Eq.(\ref{rv}) and (\ref{reff}).
The Coulomb potential in Eq.(\ref{reff}) is neglected here, 
since its role is expected to be small, compared to that from 
$\left(V_\sigma(\vct{r})-V_\omega(\vct{r})-V_\rho(\vct{r})\right)$ in Eq.(\ref{reff}). 
The value of $\left(V_\sigma(\vct{r})-V_\omega(\vct{r})-V_\rho(\vct{r})\right)$ at $r=0$
is about $-680$ MeV,  as noted below Eq.(\ref{reff}).
Eq.(\ref{nonb}) and (\ref{relb}) are accepted as the HVH theorem in asymmetric nuclear matter,
and imply the relationship between $V_\tau$ and $m^\ast_\tau$ as in Eq.(\ref{ab}),
\begin{equation}
 V_\tau=\frac{a_\tau}{m^\ast_\tau}+b_\tau, \qquad a_\tau=-(1-\delta_\tau)\frac{k^2_{{\rm F}\tau}}{2M},
 \qquad b_\tau=\left\{
 \begin{array}{ll}
\eb, &\tau=n\\[2pt]
\eb-v_c,\,&\tau=p,
\end{array}
\right. \label{mhvh1}
\end{equation}
where $\delta_\tau=0$ in the non-relativistic models, while in the relativistic models,
 $|\delta_\tau|\ll1$ being almost constant.
The values of $k^2_{{\rm F}\tau}$, which provide the values of $\rho_\tau$
in the relativistic and non-relativistic models, are determined by the two equations in Eq.(\ref{me}),
once $v_c$ is given by hand.

\begin{figure}[ht]
\centering{%
\includegraphics[scale=1]{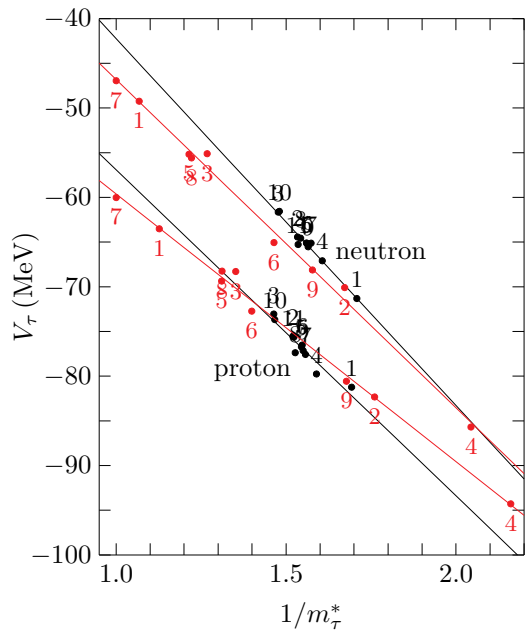}
}
 \caption{
The relationship between the effective mass and the one-body potential
for neutrons($\tau=n$) and protons ($\tau=p$) in the mean-field models for asymmetric nuclear matter.
The black circles show the values calculated by the 11 relativistic models,
while the red ones show those by the 9 non-relativistic models. Each circles are accompanied
by the number which indicates the used model specified in the text. The least-square lines are shown
for the four groups. The two black lines are obtained from the black circles
for neutrons and protons, respectively, and the red lines from the red circles.
}
\label{fig:asym_v_em}
\end{figure}

For simplicity, for both relativistic and non-relativistic models,
we take $v_c$ from the strength at $r=0$ of the Coulomb potential
for a uniformly charged sphere of radius $r_cA^{1/3}$,
\begin{equation}
 v_c=\frac{3}{2}\frac{Z\alpha}{r_cA^{1/3}}.\label{coulomb}
\end{equation}
It yields $v_c=22.144$ MeV for $^{208}$Pb with $r_c=1.350$ fm.
In employing this value, we obtain Fig.\ref{fig:asym_v_em} for the $1/m^\ast_\tau-V_\tau$
relationship corresponding to Eq.(\ref{mhvh1}).
The black circles indicate the values from the 11 relativistic models,
while the red ones from the
9 non-relativistic models. These models have been employed in Ref.\cite{kss}.
Each circle is accompanied by the number which shows the used model.
The numbering is according to Ref.\cite{kss}, as 
1.L2\cite{sw1}, 2.NLB\cite{sw1}, 3.NLC\cite{sw1}, 4.NL1\cite{nl1}, 5.NL3\cite{nl3},
6.NL-SH\cite{nlsh}, 7.NL-Z\cite{nlz}, 8.NL-S\cite{nls}, 9.NL3II\cite{nl3}, 10.TM1\cite{tm1}
and 11.FSU\cite{fsu}  for the relativistic nuclear models,
and 1.SKI\cite{sk1}, 2.SKII\cite{sk1}, 3.SKIII\cite{sk3},
4.SKIV\cite{sk3}, 5.SkM$^\ast$\cite{skm}, 6.SLy4\cite{sly4}, 7.T6\cite{st6}, 8.SGII\cite{sg2}
and 9.Ska\cite{ska} for the non-relativistic models.
The above numbering of the models will be used throughout the present paper.

In Fig.\ref{fig:asym_v_em}, the least-square lines(LSL) of these circles are drawn
by the black ones for the relativistic models,
while by the red ones for the non-relativistic models.
The LSL's for neutrons and protons are well separated from each other both in the relativistic
models and in the non-relativistic ones.
The effective mass and the one-body potential are complicated functions
of the interaction parameters whose values are different from one another
between the mean-field models. Nevertheless, as seen in Fig.\ref{fig:asym_v_em}, all the circles are almost
on their own LSL's.

On the one hand, the circle at $(m^\ast_{\tau,i}, V_{\tau,i})$ of the model $i$ is given by  
\begin{equation}
V_{\tau,i}=a_{\tau,i}/m^\ast_{\tau,i}+b_{\tau,i},\label{mhvh}
\end{equation}
according to Eq.( \ref{mhvh1}) by the HVH theorem. On the other hand, LSL satisfies
\begin{equation}
V^{\rm L}_{\tau,i}=a^{\rm L}_\tau/m^\ast_{\tau,i}+b^{\rm L}_\tau,\label{hvhline}
\end{equation}
where $a^{\rm L}_\tau$ and $b^{\rm L}_\tau$ denote the slope and intercept of LSL, respectively.
In writing the average value of Eq.(\ref{mhvh}) as $\langle V_{\tau,i}\rangle$ and that of Eq.(\ref{hvhline})
as $\langle V^{\rm L}_{\tau,i}\rangle$, they are  equal to each other by the definition of the LSL,
$\langle V_{\tau,i}\rangle=\langle V^{\rm L}_{\tau,i}\rangle$, yielding
\begin{equation}
\langle a_{\tau,i}/m^\ast_{\tau,i}\rangle
 +\langle b_{\tau,i}\rangle=a^{\rm L}_\tau\langle 1/ m^\ast_{\tau,i}\rangle+b^{\rm L}_\tau.
 \end{equation}
Hence, if the following approximation is valid,
\begin{equation}
\langle a_{\tau,i}/m^\ast_{\tau,i}\rangle
 \approx \langle a_{\tau,i}\rangle/\langle m^\ast_{\tau,i}\rangle,\quad
\langle 1/ m^\ast_{\tau,i}\rangle\approx 1/\langle m^\ast_{\tau,i}\rangle, \label{app}
\end{equation}
then we have 
\begin{equation}
\langle m^\ast_{\tau}\rangle \langle V_{\tau}\rangle\approx
  \langle a_{\tau}\rangle
+\langle b_{\tau}\rangle\langle m^\ast_{\tau}\rangle\approx
a^{\rm L}_\tau+b^{\rm L}_\tau\langle m^\ast_{\tau}\rangle \label{hvhav}
\end{equation}
by writing $\langle V_{\tau,i}\rangle=\langle V_{\tau}\rangle$,
$ \langle a_{\tau,i}\rangle=\langle a_{\tau}\rangle$, $\langle b_{\tau,i}\rangle=
\langle b_{\tau}\rangle$ and $\langle m^\ast_{\tau,i}\rangle=\langle m^\ast_{\tau}\rangle$.

\begin{table}
\begingroup
\renewcommand{\arraystretch}{1.2}
\hspace{1.5cm}%
{\setlength{\tabcolsep}{4pt}
\begin{tabular}{|c|c|c|c|c|c|c|c|c|} \hline
\multicolumn{2}{|c|}{} &
$a^{\rm L}_\tau$ &
$b^{\rm L}_\tau$ &
$\langle a_\tau \rangle$     &
$\langle b_\tau \rangle$  &
$\langle m^\ast_\tau \rangle$ &
$\langle V_\tau \rangle$ &
$\langle \rho_\tau \rangle$   \\ \hline
${\rm Rel}$   & $n$  & $-41.059$ & \hphantom{2}$-1.176$&$-38.011$&\hphantom{2}$-5.918$&$0.6426$&$-65.160$ & $0.0832$ \\ \cline{2-9}
              & $p$  & $-36.489$ & $-20.421$ & $-31.530$ & $-28.063$ & $0.6488$ & $-76.742$ &$ 0.0611$  \\ \hline
${\rm Non}$   & $n$  & $-36.743$ & $-10.069$ & $-39.927$ &\hphantom{2}$-5.840$ & $0.7524$ & $-61.219$&$0.0903$ \\ \cline{2-9}
              & $p$  & $-29.960$ & $-29.656$ & $-31.168$ & $-27.984$ & $0.7217$ & $-73.256$&$0.0623$  \\ \hline 
\end{tabular}
}
\endgroup
\caption{The values of the gradient($a^{\rm L}_\tau$) and the intercept($b^{\rm L}_\tau$) of the least-square lines
in units of MeV for the relationship between $1/m^\ast_\tau$ and $V_\tau$ in Fig.\ref{fig:asym_v_em}.
The average values of the coefficients in Eq.(\ref{mhvh}) are also listed as $\langle a_\tau \rangle$ and $\langle b_\tau \rangle$,
together with  those of the effective masses, the strengths of the one-body potentials\,(MeV)
and the nuclear matter densities\,(\,fm$^{-3}$\,) in the relativistic\,(Rel) and non-relativistic\,(Non) models.
 The notations of $n$ and $p$ indicate that the values in the corresponding rows
are for neutrons and protons, respectively.
For details, see the text.}
\label{table_2}
\end{table}

In Table 2 are listed the values of the slope $a^{\rm L}_\tau$ and intercept $b^{\rm L}_\tau$
of the LSL. The average values of $a_{\tau,i}$ and $b_{\tau,i}$ calculated by each model are 
tabulated as $\langle a_\tau \rangle$ and $\langle b_\tau \rangle$.
The average values of the effective mass $\langle m^\ast_\tau \rangle$ and
of the strengths of the one-body potentials $\langle V_\tau \rangle$ are also listed,
together with the average values of $\rho_{\tau,i}$ as $\langle \rho_\tau \rangle$.

The difference between the values of $\langle a_\tau\rangle$ in the relativistic and non-relativistic models
is related to those of $\langle \rho_\tau \rangle$ through the Fermi momentum.
The values of $\langle b_\tau\rangle$ are almost the same between the two frameworks,
since $\langle b_\tau\rangle$
satisfies the relationship as $\langle \eb \rangle=\langle b_n \rangle$
and $v_c=\langle b_n \rangle-\langle b_p\rangle$,
where $\langle \eb \rangle$ denote the average value of ${\eb}_{,i}$,
according to Eq.(\ref{mhvh1}).

The values of $a^{\rm L}_\tau$ and $b^{\rm L}_\tau$ depend on the distributions
of the points ($m^\ast_{\tau,i},V_{\tau,i}$) and have no simple relationship to $\langle \rho_\tau \rangle$,
$\langle \eb \rangle$ and $v_c$. They, however, are implicitly constrained by the HVH
theorem through Eq.(\ref{hvhav}). Since the values of $v_c$ and $\langle \eb \rangle$ 
are almost the same in the relativistic and non-relativistic models, Eq.(\ref{hvhav}) provides
the relationship between the effective mass, the strength of the one-body potential and the nucleon density.
This fact implies that the LSL coefficients $a^{\rm L}_\tau$ and $b^{\rm L}_\tau$ are dominated by
$\langle \rho_\tau \rangle$ implicitly.

Eq.(\ref{hvhav}) is rewritten as $a^{\rm L}_\tau-\langle a_{\tau}\rangle
\approx \bigl(\langle b_{\tau}\rangle-b^{\rm L}_\tau\bigr)\langle m^\ast_{\tau}\rangle$,
which provides the relationship as
$a^{\rm L}_\tau \gtrsim\langle a_{\tau}\rangle$ for $b^{\rm L}_\tau\lesssim\langle b_{\tau}\rangle$,
and $a^{\rm L}_\tau\lesssim\langle a_{\tau}\rangle$ for $ b^{\rm L}_\tau\gtrsim
 \langle b_{\tau}\rangle$.
The non-relativistic models obey the first case, while the relativistic ones the second case.
The value of $|b^{\rm L}_n|$  is made much smaller by the small $\langle m^\ast_n\rangle$
in the relativistic models, compared to that in the non-relativistic one.

\begin{table}
\begingroup
\renewcommand{\arraystretch}{1.2}
\hspace{3.5cm}%
{\setlength{\tabcolsep}{4pt}
\begin{tabular}{|c|c|c|c|c|} \hline
\multicolumn{2}{|c|}{}     &
$\langle m^\ast_\tau \rangle\langle V_\tau \rangle$ &
$a^{\rm L}_\tau +b^{\rm L}_\tau\langle m^\ast_{\tau}\rangle $ &
$\langle a_{\tau} \rangle + \langle b_{\tau}\rangle\langle m^\ast_{\tau}\rangle$
 \\ \hline
${\rm Rel}$   & $n$  & $-41.872$ & $-41.815$ & $-41.814$ \\ \cline{2-5}
              & $p$  & $-49.790$ & $-49.738$ & $-49.737$ \\ \hline
${\rm Non}$   & $n$  & $-46.061$ & $-44.319$ & $-44.321$ \\ \cline{2-5}
              & $p$  & $-52.869$ & $-51.363$ & $-51.364$ \\ \hline 
\end{tabular}
}
\endgroup
\caption{The product of the mean values of the effective mass $\langle m^\ast_\tau\rangle$
 and the one-body potential $\langle V_\tau\rangle$. The numbers are given in units of MeV.
For details, see the text.}
\label{table_3}
\end{table}

In Table 3, the value of each term in Eq.(\ref{hvhav}) is listed. It shows that
the values of $\langle m^\ast_{\tau}\rangle \langle V_{\tau}\rangle$ 
are a little larger than those of $\langle a_{\tau}\rangle
+\langle b_{\tau}\rangle\langle m^\ast_{\tau}\rangle$
and $a^{\rm L}_\tau+b^{\rm L}_\tau\langle m^\ast_{\tau}\rangle$ in the non-relativistic models,
because the approximations in Eq.(\ref{app}) are a little worse in the non-relativistic models
than in the relativistic models.
This difference, however, is not essential for the present discussions on $\delta R$.

In finite nuclei, Eq.(\ref{ra}) indicates that the radius depends on
$\bigl(-\langle m^\ast_\tau \rangle\langle V_\tau \rangle\bigr)^{-1/4}$.   
Table 3 implies a possibility that $R_n$ is larger in the relativistic models than
in the non-relativistic ones, if the same tendency maintains in finite nuclei.
In the mean-field models for finite nuclei, however, the effective mass and one-body potential
may have complicated coordinate dependences.
In order to confirm the above implications for finite nuclei, 
we need a way to extract from them the values of the effective mass and
the strength of the one-body potentials
which are appropriate for the use in Eq.(\ref{ra}).
Moreover, it is desirable to explore whether or not they are constrained
by the HVH theorem as in asymmetric nuclear matter.
Although we do not have for finite nuclei an equation like Eq.(\ref{me})
to yield the HVH constraint in asymmetric nuclear matter,
Eq.(\ref{hvhav}) may be helpful for understanding roles of the HVH theorem in finite nuclei. 
Bearing these facts in mind, we proceed discussions for finite nuclei from the next section.

\section{Simplification of the mean-field models}\label{simp}

One of the ways to find a common structure of the models is to simplify them without loosing
their main characteristics.  By defining the effective mass and the one-body potential
by such a way, we may find their relationships to $R$ and a restriction
between them like the HVH lines which are hidden in the complexity
of the calculated results of the mean-field models for finite nuclei.

In this section, we will analyze the structure of the relativistic and non-relativistic models
by simplifying their descriptions as much as possible.
As mentioned in \S\ref{relation}, $R_\tau$ may be a functional
of $V_\tau(r), M^\ast_\tau(r), V_c(r)$ and the spin-orbit potential, $V_{\ell s,\tau}(r)$,
but among them, it is expected for $V_c(r)$
and $V_{\ell s,\tau}(r)$ to play a minor role in the difference between $\delta R$'s
in the two frameworks, $\delta R[V_\tau, M^\ast_\tau,
V_c,V_{\ell s,\tau}]\approx \delta R[V_\tau, M^\ast_\tau]$.
Using these facts as a guide, let us express approximately all the Hamiltonians
in the both frameworks, using the same basis.

\subsection{Nuclear potential and effective mass}\label{ws}

The fundamental properties of nuclei are well described
with  the WS potential\cite{bm}, and its structure is clear for the present purpose to
discuss $\delta R$, as in Eq.(\ref{rra}).
Hence, we approximate the mean-field potential, $V_\tau(r)$, and the effective mass,
$M^\ast_\tau(r)$, in both frameworks by using the WS type function,
\begin{equation}
f_\tau(r)=f_\tau(r,R_\tau,a_\tau)=\frac{1}{1+\exp\left((r-R_\tau)/a_\tau \right)},
 \end{equation}
that is,
\begin{align}
V_\tau(r)
&\approx
V_{{\rm ws},\tau}f(r,R_{{\rm ws},\tau},a_{{\rm ws},\tau}), \label{wsp}\\[4pt]
m^\ast_\tau(r)
&\approx
\left\{
\begin{array}{ll}
  1+(m^\ast_{{\rm ws},\tau}-1) f(r,R^\ast_{{\rm ws},\tau},a^\ast_{{\rm ws},\tau})
   -\dfrac{V_c(r)}{2M}\dfrac{1+\tau_3}{2},  & ({\rm Rel}),\\[12pt] 
   1+(m^\ast_{{\rm ws},\tau}-1) f(r,R^\ast_{{\rm ws},\tau},a^\ast_{{\rm ws},\tau}),
    & ({\rm Non}),
 \end{array}
\right. \label{nm}
\end{align}
where $m^\ast_\tau(r)=M^\ast_\tau(r)/M$ is defined, and Rel and Non indicate the relativistic and
non-relativistic models, respectively.
The three-parameters in the right-hand sides of the above equations are determined by minimizing,
for example, for $V_\tau(r)$, the following quantity with respect to $V_{{\rm ws},\tau},R_{{\rm ws},\tau}$,
and $a_{{\rm ws},\tau}$,
\begin{equation}
 \int^\infty_0 dr\,r^2\Bigl( V_\tau(r)-V_{{\rm ws},\tau}f(r,R_{{\rm ws},\tau},a_{{\rm ws},\tau}) \Bigr)^2.
  \label{min}
\end{equation}
Here, the volume integral has been chosen in order to minimize
the above deviation, since both $V_\tau(r)$ and $V_{\rm ws,\tau}f(r,R_{{\rm ws},\tau},a_{{\rm ws},\tau})$
are expected to have a similar shape
to that of the nuclear density whose volume integral value is constrained by the nucleon number.
In deriving Eq.(\ref{ra}), we have used $n=0$ in Eq.(\ref{wmin}),
since there is not such a constraint on the HO potential,
but since it is important to keep
a similarity of the wave functions in the HO and WS potential.

\subsection{Coulomb potential}\label{cp}

In the relativistic models, the Coulomb energy is calculated by taking into account
the only direct term of the interaction in the same way as for other interactions,
while the exchange term is also included in the non-relativistic models.
The Coulomb interaction, however, plays a minor role in the present purpose on $\delta R$,
so that we simply express it by that of the uniform charge distribution
with the radius, $R_{\rm coul}$,
\begin{equation}
 V_c(r)\approx V_{\rm sph}(r), \qquad V_{\rm sph}(r)=\left\{
 \begin{array}{ll}
 \dfrac{Z\alpha}{2R_{\rm coul}}\left(3-\dfrac{r^2}{R^2_{\rm coul}}\right)\,, & r < R_{\rm coul},\\[12pt]
 \dfrac{Z\alpha}{r}, & r >R_{\rm coul}.
   \end{array}
   \right.\label{cou}
\end{equation}
The radius $R_{\rm coul}$ is determined by minimizing the deviation:
\[
 \int^\infty_0 dr\,r^2\Bigl( V_c(r)-V_{\rm sph}(r) \Bigr)^2.
\]
The value of $R_{\rm coul}$ of the each model will be shown later.

\subsection{Spin-orbit potential}

We express the spin-orbit potential in the form:
\begin{equation}
V_{\ell s,\tau}(r)=V_{\ell s,\tau}\frac{1}{r}\frac{df_\tau(r)}{dr}
 \vct{\ell}\!\cdot\!\vct{\sigma}.
\end{equation}
In the relativistic models, it is written from Eq.(\ref{rse}) as
\begin{align}
V_{\ell s,\tau}(r)=\frac{1}{r}\frac{d}{dr}\frac{1}{2M^\ast_\tau(r)}
\vct{\ell}\!\cdot\!\vct{\sigma}
&=-\frac{1}{2M}\frac{1}{m^{\ast 2}_\tau(r)}\frac{1}{r}
 \frac{dm^\ast_\tau(r)}{dr}\vct{\ell}\!\cdot\!\vct{\sigma}\nonumber\\[2pt]
 &\approx\frac{1-m^\ast_{{\rm ws},\tau}}{2M}\frac{1}{m^{\ast 2}_\tau(r)}\frac{1}{r}
 \frac{d}{dr}f(r,R^\ast_{{\rm ws},\tau},a^\ast_{{\rm ws},\tau})\vct{\ell}\!\cdot\!\vct{\sigma},
\end{align}
with neglecting $V_c(r)$ in Eq.(\ref{nm}). In the calculations, the further 
approximation has been used, as 
$m^\ast_\tau(r)\approx m^\ast_\tau(R^\ast_{{\rm ws},\tau})=(1+m^\ast_{{\rm ws},\tau})/2$.

In the non-relativistic models, the spin-orbit potential of Eq.(\ref{nls}) is approximated as
\begin{equation}
 V_{\ell s,\tau}(r)=\frac{W_0}{2r}\frac{d}{dr}\Bigl(\rho(r)+\rho_\tau(r)\Bigr)\vct{\ell}\!\cdot\!\vct{\sigma}
\approx \frac{W_0\rho_0}{2}\left(1+\frac{N_\tau}{A}\right)\frac{1}{r}\frac{d}{dr}f(r,R_{{\rm den}},
a_{{\rm den}})\,\vct{\ell}\!\cdot\!\vct{\sigma},
\end{equation}
where the value of $W_0$ is fixed at 120\,MeV\,fm$^5$ and
we have written the nuclear density as
\[
 \rho(r)=\rho_p(r)+\rho_n(r)\approx\rho_0f(r,R_{{\rm den}},a_{{\rm den}})\,,\quad
\]
with
\[
4\pi\rho_0\int^\infty_0dr\,r^2f(r,R_{{\rm den}},a_{{\rm den}})=A.
\]
The details of the calculation of the nuclear density will be mentioned in \S\ref{pndis}.

In fact, the spin-orbit potentials are expected not to play an important role
in understanding the difference of $\delta R$ between the relativistic and non-relativistic models,
since their strengths are similar and the isospin-dependences are small, in addition to the reason
mentioned before.

\subsection{A few examples}

Before summarizing the results of the present section, let us compare
$ V_\tau(r),\,  \rho_\tau(r)$ and $m^\ast_\tau(r)$ from the exact mean-field calculations
with those of the corresponding simplified Hamiltonian, by taking a few examples.
After minimizing Eq.(\ref{min}), the only values of $V_{{\rm ws}, n}$ for the relativistic WS
potentials have been multiplied by 0.99, so as to reproduce well the values of $R_n$
in the exact relativistic mean-field calculations.
This factor makes the mean value of $R_n$ from the WS
potential smaller by about $0.015$ fm.

\begin{figure}[ht]
\begin{minipage}[t]{7.2cm}
\includegraphics[scale=0.9]{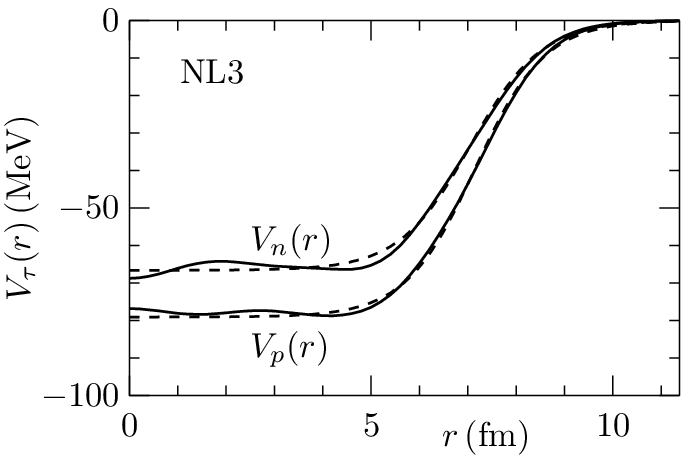}%

\vspace{0.5cm}%
\includegraphics[scale=0.9]{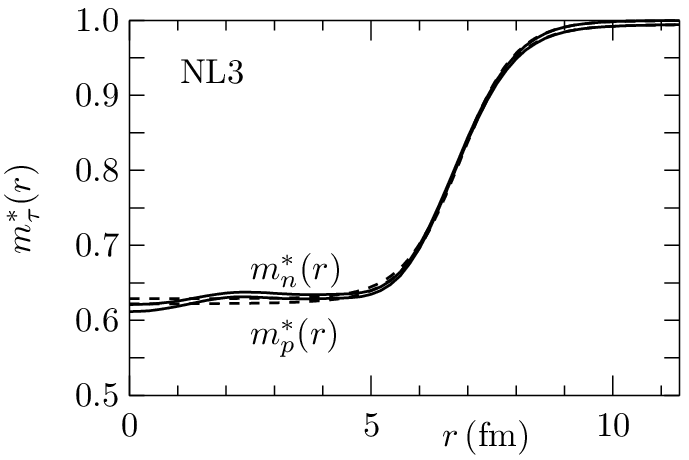}%

\vspace{0.5cm}%
\includegraphics[scale=0.9]{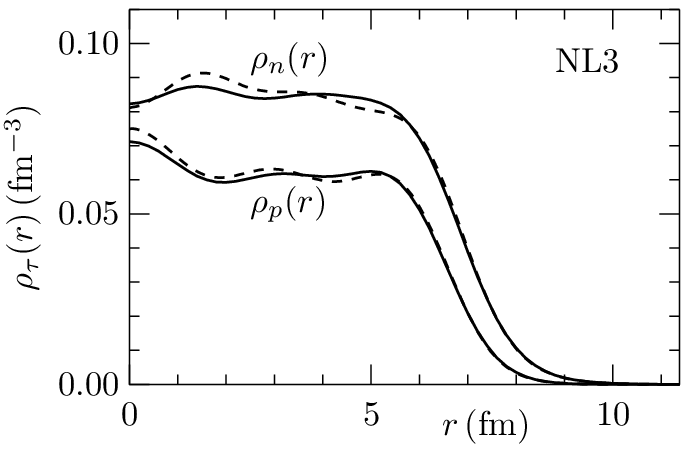}%
\caption{The one-body potentials, the effective masses and the densities
for neutrons($\tau=n$) and protons($\tau=p$) in $^{208}$Pb.
The solid curves are obtained by the relativistic mean-field model with NL3,
while the dashed ones by its simplified Hamiltonian.}
\label{fig:NL3}
\end{minipage}
\hspace{0.6cm}
\begin{minipage}[t]{7.2cm}
\includegraphics[scale=0.9]{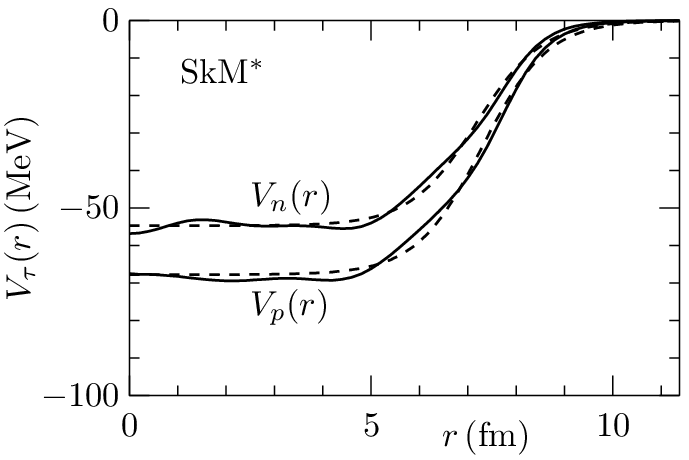}%

\vspace{0.5cm}%
\includegraphics[scale=0.9]{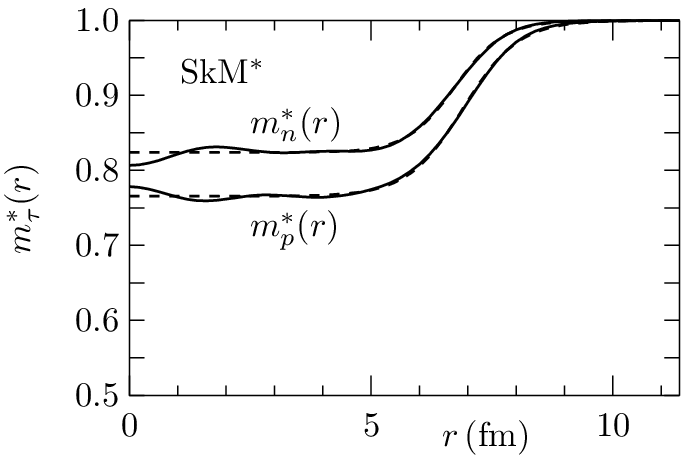}%

\vspace{0.5cm}%
\includegraphics[scale=0.9]{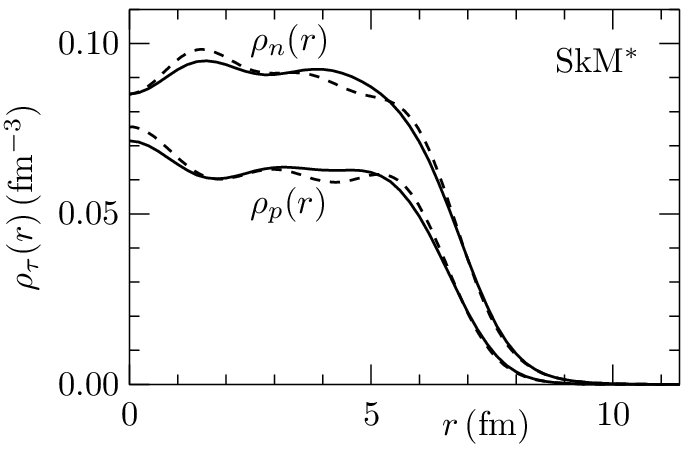}%
\caption{The one-body potentials, the effective masses and the densities
for neutrons($\tau=n$) and protons($\tau=p$) in $^{208}$Pb.
The solid curves are obtained by the non-relativistic mean-field model with SkM$^\ast$,
while the dashed ones by its simplified Hamiltonian.}
\label{fig:SkMm}
\end{minipage}
\end{figure}

Fig.\ref{fig:NL3} shows the results of the one-body potentials, the effective masses and the neutron
and the proton densities for $^{208}$Pb calculated with NL3.
The solid curves are obtained by the full calculations and the dashed ones by the simplified
Hamiltonians. All other relativistic models yield similar results.
In non-relativistic models, we show the results for SkM$^\ast$ in Fig.\ref{fig:SkMm}.
These results of SKM$^\ast$ are similar to those of other models except for the effective mass in SLy4.
In SLy4, the coordinate dependences of the effective mass are similar to those in Fig.\ref{fig:SkMm},
but the relation of the magnitude between the $m^\ast_{{\rm ws},p}$ and $m^\ast_{{\rm ws},n}$ is opposite to
that in other non-relativistic models. 
It is seen that all the results by simplified versions well reproduce
the corresponding ones obtained by the full calculations.

\subsection{Results by the simplified models}\label{result}

Table 4 shows the root msr's of the point neutron distributions in $^{48}$Ca and $^{208}$Pb
determined in Ref.\cite{kss}.
Those of the point proton distributions obtained in a similar way are also listed.
The errors in the parentheses are given by taking into account the experimental error
and the standard deviation of the LSL.
Since both relativistic and non-relativistic models employ the experimental values of the msr's
of the nuclear charge distributions as an input, the values of $R_p$ in the two frameworks
are almost equal to each other,
while the values of $R_n$ are larger by about 0.1 fm in the relativistic models than
in the non-relativistic ones in both $^{48}$Ca and $^{208}$Pb.
The purpose of the present paper is to understand this difference between $R_n$'s.

\begin{table}
\begingroup
\renewcommand{\arraystretch}{1.2}
\hspace{3.6cm}%
{\setlength{\tabcolsep}{4pt}
\begin{tabular}{|c|c|c|c|c|} \hline
\multicolumn{2}{|c|}{} &
$R_n$ & $R_p$ & $\delta R$  \\ \hline
           & Rel & $3.597(0.021)$ & $3.378(0.005)$  & $0.220(0.026)$   \\
$^{48}$Ca  & Non & $3.492(0.028)$ & $3.372(0.009)$  & $0.121(0.036)$   \\ \hline
%%%
           & Rel & $5.728(0.057)$ & $5.454(0.013)$  & $0.275(0.070)$   \\
$^{208}$Pb & Non & $5.609(0.054)$ & $5.447(0.014)$  & $0.162(0.068)$   \\ \hline
\end{tabular}
}
\endgroup
\caption{The results of the least squares analysis in Ref.\cite{kss}. The numbers in the parentheses denote
the error which is obtained taking account of the experimental error and the standard deviation of the
calculated values from the least-square line.  All the numbers are given in units of fm. For details, see the text.}
\label{table_4}
\end{table}

We note that the new data from JLab have been reported in Ref.\cite{adh}, according to the parity
violating electron scattering experiment. It provides $\delta R$ in $^{208}$Pb to be
$0.283\pm 0.071$ fm. 
This is almost the same as the value of $\delta R$ in Table 3 in the relativistic models,
and is not incompatible with the non-relativistic one in taking into account their errors.
The analysis of the JLab data\cite{adh} is model-dependent as in Ref.\cite{kss},
and Ref.\cite{naz} has obtained $\delta R=0.19\pm0.02$ fm from the JLab data on the basis of
the different model-analysis.

\begin{figure}[ht]
\begin{center}
\includegraphics[scale=1]{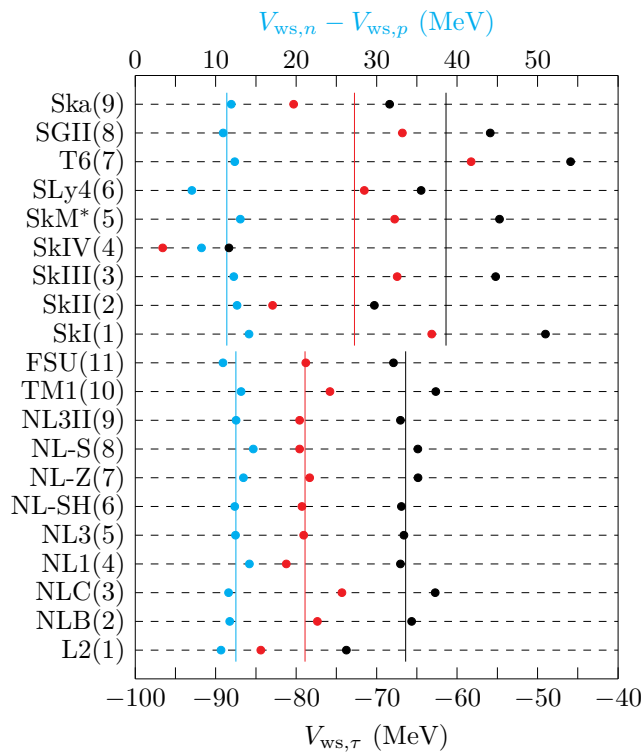}%
\caption{The strength of the one-body potentials for neutrons($\tau=n$) and protons($\tau=p$)
 in $^{208}$Pb.
On the left-hand side, the relativistic and non-relativistic models are indicated.
The black circles represent the strengths of the neutron potentials, the red ones
those of the proton potentials, and the blue ones the differences between their two strengths.
The vertical lines stand for the average values of the corresponding circles
in the relativistic and non-relativistic models, separately.
The scale of the bottom side is for the strength of the potentials,
while that of the top side for the difference between the strengths of neutron
and proton potentials.
}
\label{fig:Vws_force}
\end{center}
\end{figure}

Let us summarize the results in the present section for $^{208}$Pb.
Fig.\ref{fig:Vws_force} shows the values of $V_{{\rm ws},\tau}$ in Eq.(\ref{wsp}) for the relativistic and
non-relativistic models. 
The strengths of the neutron potentials are shown by the black circles,
and those of the proton ones by the red circles.
It is seen that the non-relativistic ones are distributed over a wide range, as expected,
in contrast to those of relativistic models. The straight vertical lines show their average values. 
The difference between $V_{{\rm ws},n}-V_{{\rm ws},p}$, however, is almost equal independently
of the models, as shown by the blue circles
and the straight lines indicating their average values.
Thus, the difference is only a little larger in the relativistic models than in the
non-relativistic ones. This fact implies that the difference between $\delta R$'s
in the two frameworks may not be due to the symmetry potentials only.

\begin{figure}[ht]
\begin{minipage}[t]{7.2cm}
\includegraphics[scale=0.9]{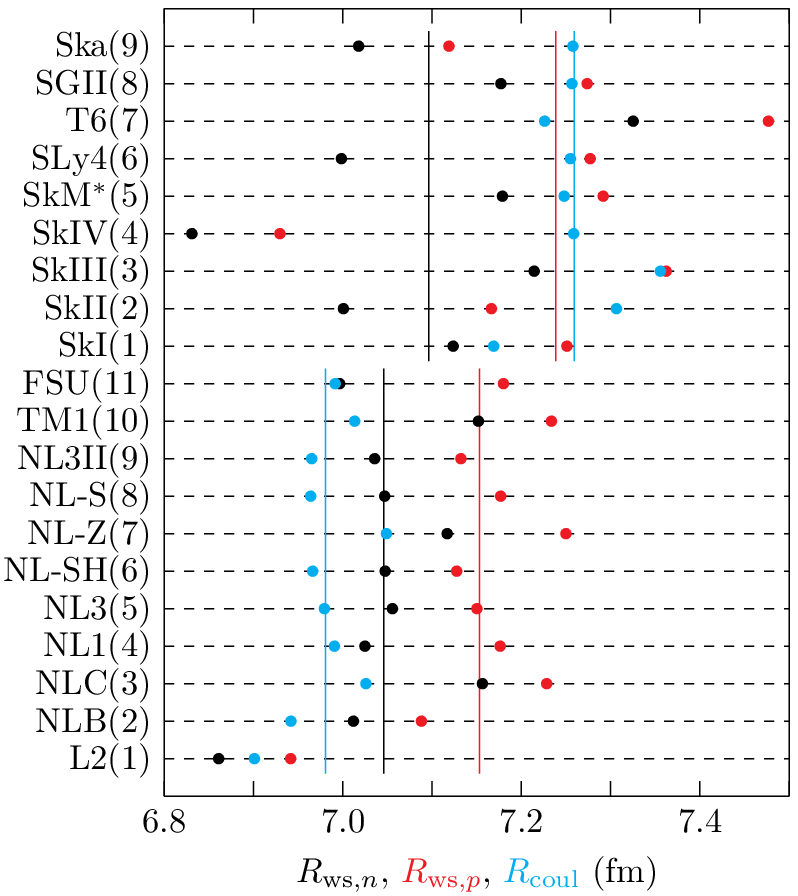}%
\caption{The radius parameters of the Woods-Saxon potentials for neutrons($\tau=n$) and protons($\tau=p$)
in $^{208}$Pb obtained from the relativistic and  non-relativistic mean-field models. 
On the left-hand side, the used relativistic and non-relativistic models are indicated.
The black circles represent the values of the radius parameter $R_{{\rm ws},n}$,
while the red ones those of $R_{{\rm ws},p}$.
The blue ones stand for the values of the radius parameters of the Coulomb potentials.  
The vertical lines indicate the average values of the corresponding circles 
in the relativistic and non-relativistic models, separately.
}
\label{fig:Rws_force}
\end{minipage}\hspace{0.8cm}%
\begin{minipage}[t]{7.2cm}
\includegraphics[scale=0.9]{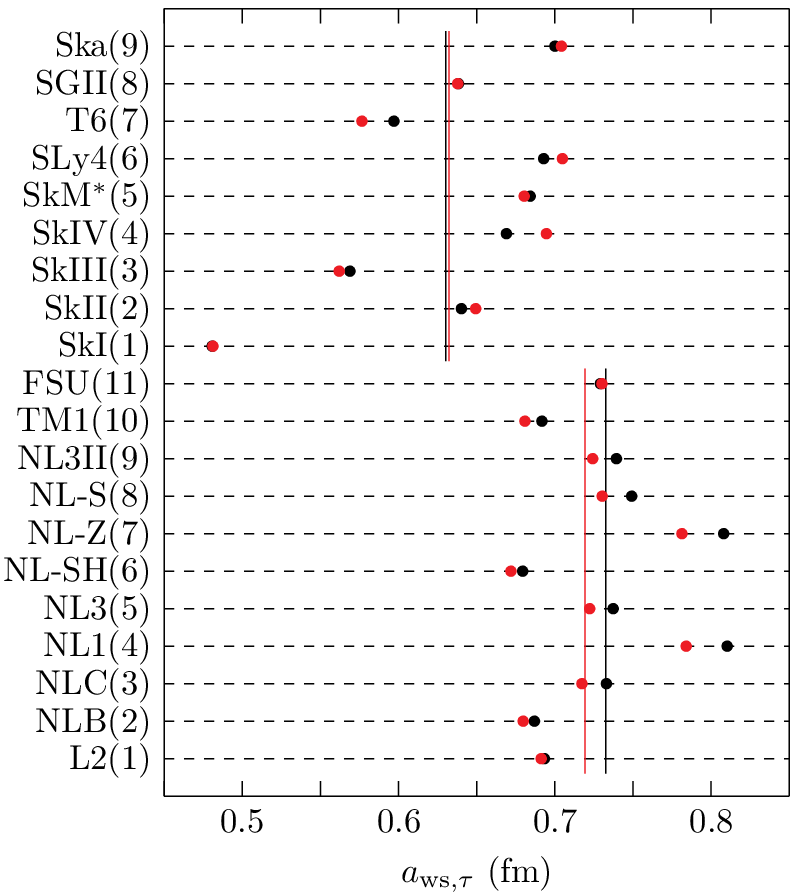}%
\caption{The diffuseness parameters of the Woods-Saxon potentials for neutrons($\tau=n$)
and protons($\tau=p$) in $^{208}$Pb obtained from the relativistic and  non-relativistic
mean-field models. 
On the left-hand side, the used relativistic and non-relativistic models are indicated.
The black circles represent the values of the diffuseness parameter $a_{{\rm ws},n}$,
while the red ones those of $a_{{\rm ws},p}$.
The vertical lines indicate the average values of the corresponding circles, 
in the relativistic and non-relativistic models, separately.}
\label{fig:aws_force}
\end{minipage}
\end{figure}

Fig.\ref{fig:Rws_force} shows the values of $R_{{\rm ws},\tau}$ in a similar way
as in Fig.\ref{fig:Vws_force}.
The black and red circles for the non-relativistic models
are again distributed over a wide region, compared to those of the relativistic ones,
although their regions are overlapped.
The solid lines express the mean values of the corresponding circles.
It is seen that the value of the difference,
$\langle R_{{\rm ws},p}\rangle-\langle R_{{\rm ws},n}\rangle$,
in the relativistic models is rather smaller than that in the non-relativistic models as in Table 5.
Thus, the spread of the values of $R_{{\rm ws},\tau}$ in the non-relativistic models
does not seem to cause the difference between $\delta R$'s in the two frameworks.
The values of $R_{\rm coul}$ are indicated by the blue circles for reference.

Fig.\ref{fig:aws_force} shows the values of $a_{{\rm ws},\tau}$. The straight lines stand for their
average values. Those of the relativistic and non-relativistic models are distributed
similarly over a wide region, but are small, compared to $R_{{\rm ws},\tau}$,
as $(a_{{\rm ws},\tau}/R_{{\rm ws},\tau})^2\approx0.01$.
The difference between $\delta R$'s in the relativistic and non-relativistic models
may not be due to these distributions of $a_{{\rm ws},\tau}$.

\begin{figure}[ht]
\begin{center}
\includegraphics[scale=1]{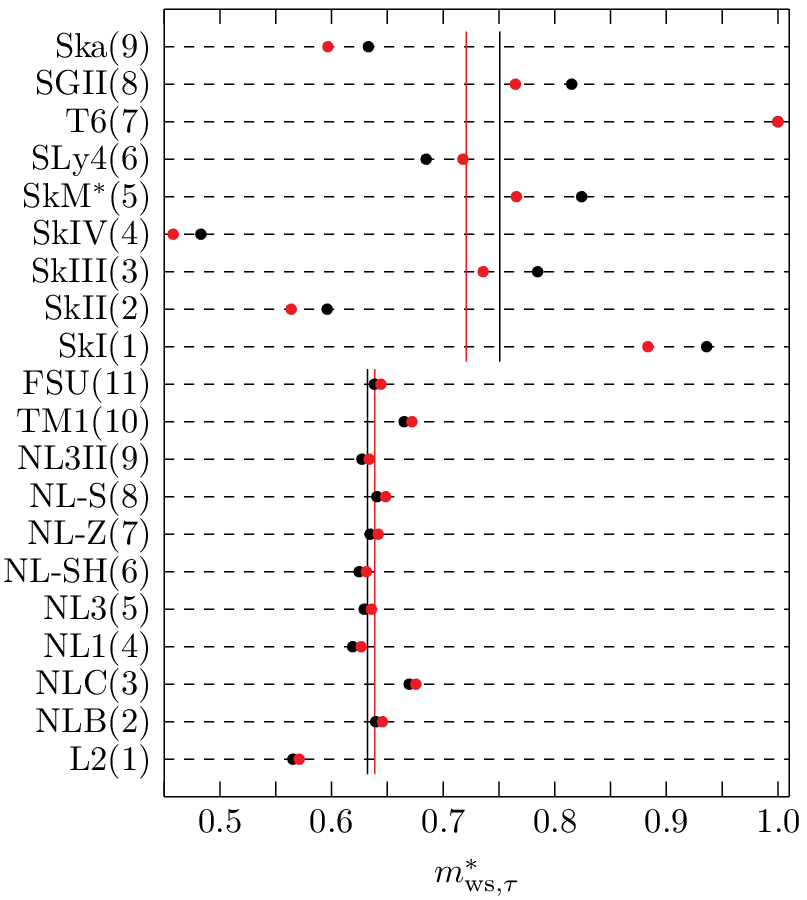}%
\caption{
The effective mass of neutrons($\tau=n$) and protons($\tau=p$) in $^{208}$Pb
obtained from the relativistic and  non-relativistic mean field models. 
On the left-hand side, the used relativistic and non-relativistic models are indicated.
The black circles represent the values of the effective masses of neutrons,
while the red ones those of protons. In T6, the value for neutrons is
the same as that for protons.
The vertical lines indicate the average values of the corresponding circles, 
in the relativistic and non-relativistic models, separately.
 }
\label{fig:mws_force}
\end{center}
\end{figure}

In Fig.\ref{fig:mws_force} are shown the values of $m^\ast_{{\rm ws},\tau}$.
The black circles represent
those of the neutrons, while the red circles the protons.
The straight lines indicate their
average values, which are in the relativistic models $\langle m^\ast_{{\rm ws},n}\rangle\approx 0.6321$ and
$\langle m^\ast_{{\rm ws},p}\rangle\approx 0.6387$, and in the non-relativistic models
$\langle m^\ast_{{\rm ws},n}\rangle\approx 0.7508$ and  $\langle m^\ast_{{\rm ws},p}\rangle\approx 0.7207$.
As seen in the figure, the circles of the relativistic models are almost at the same value
and the ratio, $\langle m^\ast_{{\rm ws},p}\rangle/\langle m^\ast_{{\rm ws},n}\rangle$, is about 1.010,
while those of the non-relativistic models are spread out, as in nuclear matter, 
but the ratio in  each model is almost the same
and is on average $\langle m^\ast_{{\rm ws},p}\rangle/\langle m^\ast_{{\rm ws},n}\rangle\approx 0.9601$.

Table 5 lists the mean values of the WS parameters, the strengths of the one-body potentials
and the effective masses in the present simplified models for the relativistic and non-relativistic
mean-field ones, respectively. It should be noticed that the values of $\langle m^\ast_{{\rm ws},\tau}\rangle$
and $\langle V_{{\rm ws},\tau}\rangle$ are almost the same as those of $\langle m^\ast_\tau\rangle$
and $\langle V_\tau\rangle$ in Table 2.

\begin{table}
\begingroup
\renewcommand{\arraystretch}{1.2}
\hspace{1cm}%
{\setlength{\tabcolsep}{4pt}
\begin{tabular}{|c|c|c|c|c|c|c|c|c|} \hline
&
$\langle R_{{\rm ws},n}\rangle$ &
$\langle R_{{\rm ws},p}\rangle$ &
$\langle a_{{\rm ws},n}\rangle$ & 
$\langle a_{{\rm ws},p}\rangle$ &
$\langle V_{{\rm ws},n}\rangle$ &
$\langle V_{{\rm ws},p}\rangle$ &
$\langle m^\ast_{{\rm ws},n}\rangle$ &
$\langle m^\ast_{{\rm ws},p}\rangle$ \\ \hline 
Rel  & $7.046$ & $7.153$ & $0.733$  & $0.719$  &$-66.376$ &$-78.888$&$0.6321$&$0.6387$ \\ \cline{1-9}
Non & $7.096$ & $7.239$ & $0.630$  & $0.632$&$-61.368$ &$-72.759$&$0.7508$&$0.7207$\\ \cline{1-9}
\end{tabular}
}
\endgroup
\caption{ The mean values of the Woods-Saxon parameters, the strengths of the one-body potentials
 and the effective masses in the relativistic and  non-relativistic models.
 The numbers of $\langle R_{{\rm ws},\tau}\rangle$ and $\langle a_{{\rm ws}, \tau}\rangle$
 are given in units of fm, and those of $\langle V_{{\rm ws},\tau}\rangle$ in units of MeV.
 For details, see the text.}
\label{table_5}
\end{table}

\subsection{The proton and neutron distributions}\label{pndis}

It may be useful to see directly how the difference between $\delta R$'s is caused
in terms of the neutron and proton densities.
We approximate the neutron and proton distributions, $\rho_\tau(r)$, in the mean-field models 
by the Fermi-type function which is widely employed\cite{bm, vries}.
The approximation is performed for
\begin{equation}
 \rho_{{\rm ws},\tau}(r)\approx \rho_{0,\tau}f_\tau(r,R_{{\rm den},\tau}, a_{{\rm den},\tau}),\label{fermi}
\end{equation}
in the same minimization method as in Eq.(\ref{min}).
Since the obtained function satisfies the normalization, $\int d^3r\rho_{{\rm ws},\tau}(r)=N_\tau $
with a small error about $0.5\%$, we slightly correct $R_{{\rm den},\tau}$ to satisfy the 
normalization.
We note that the correction by $\rho_{0,\tau}$ instead of $R_{{\rm den},\tau}$ yields
almost the same values as those which will be seen in Tables 6 and 7.
The minimization under the constraint on the nucleon-number also yields similar results.

The msr, $R^2_\tau$, by Eq.(\ref{fermi}) is given as\cite{bm},
\begin{equation}
R^2_\tau \approx \frac{3}{5}\left(\frac{3N_\tau}{4\pi\rho_{0,\tau}}\right)^{2/3}
 +\pi^2 a^2_{{\rm den},\tau},\label{approxr}
 \end{equation}
which provides the relationship between $R^2_n$ and $R^2_p$ as
\begin{equation}
R^2_n=\frac{R^2_p-\pi^2a^2_{{\rm den},p}}{(1-\epsilon)^{2/3}}+\pi^2a^2_{{\rm den},n},
 \quad \epsilon=1-\frac{Z}{N}\frac{ \rho_{0,n}}{ \rho_{0,p}}. \label{rn}
\end{equation}
When keeping order up to $O(\epsilon)$, the difference, $\delta R=R_n-R_p$, is written as
\begin{equation}
\delta R \approx \frac{\epsilon}{3}R_p+\frac{\pi^2}{2}\frac{a^2_{{\rm den},n}-a^2_{{\rm den},p}}{R_p}.
\label{approxd}
\end{equation}

\begin{table}
\begingroup
\renewcommand{\arraystretch}{1.2}
{\setlength{\tabcolsep}{4pt}
\hspace{4cm}
\begin{tabular}{|c|c|c|c|c|c|} \hline
\multicolumn{2}{|c|}{}     &
$\rho_{0,\tau}$ &
$a_{{\rm den},\tau}$   &
$R_{{\rm den},\tau}$  &
$\epsilon$  \\ \hline 
Rel   & $n$  & $0.0860$ & $0.553$& $6.903$&$0.1044$  \\ \cline{2-5}
      & $p$  & $0.0625$ & $0.454$& $6.692$&          \\ \hline
Non   & $n$  & $0.0911$ & $0.554$ &$6.766$ &$0.0556$ \\ \cline{2-5}
      & $p$  & $0.0628$ & $0.475$ &$6.672$ &         \\ \hline 
\end{tabular}
}
\endgroup
\caption{The mean values of the parameters for the Fermi-type neutron(n) and proton(p) distributions.
They are obtained by approximating the densities in the relativistic and non-relativistic
mean-field models for $^{208}$Pb. The values of $\rho_{0,\tau}$ are given in units of fm$^{-3}$
and those of $a_{{\rm den},\tau}$ and $R_{{\rm den},\tau}$ are in fm. For the definition of $\epsilon$, see the text.}
\label{table_6}
\end{table}

\begin{table}
\begingroup
\renewcommand{\arraystretch}{1.2}
{\setlength{\tabcolsep}{4pt}
\hspace{4.5cm}
\begin{tabular}{|c|c|c|c|c|} \hline
\multicolumn{2}{|c|}{}       &
$R_n$  &
$R_p$  &
$\delta R$  \\ \hline
Rel   & MF                & 5.749 & 5.466 & 0.283 \\ \cline{2-5}
      &WS                 & 5.740 & 5.457 & 0.283 \\ \cline{2-5}
      &Eq.(\ref{approxr}) & 5.728 & 5.451 & 0.277 \\ \hline
Non   &MF                 & 5.617 & 5.455 & 0.161 \\ \cline{2-5}
      &WS                 & 5.621 & 5.462 & 0.159 \\ \cline{2-5}
      &Eq.(\ref{approxr}) &5.629&5.460&0.169   \\ \hline 
\end{tabular}
}
\endgroup
\caption{The mean values of the root msr's of the neutron($R_n$) and proton($R_p$) distributions 
and their difference $(\delta R=R_n-R_p)$ obtained by the three approximations for $^{208}$Pb. 
MF indicates the mean values in the relativistic and non-relativistic mean-field models,
while WS the ones obtained by approximating the mean-field potentials by Woods-Saxon potentials.
Eq.(\ref{approxr}) stands for the equation number in the text used for the calculations of $R_\tau$
and $\delta R$ given in its row. All the values are listed in units of fm. For details, see the text.}
\label{table_7}
\end{table}

Table 6 lists the average values of the parameters for the Fermi-type densities in Eq.(\ref{fermi})
in the present relativistic and non-relativistic models.
Table 7 shows the values of Eq.(\ref{approxr}) using the results in Table 6.
The average values of $R_\tau$ and $\delta R$ in the mean-field models and their simplified
versions also are listed in the rows named MF and WS, respectively.
It is seen that the values in Eq.(\ref{approxr}) and in WS almost reproduce the results
of the mean-field models.

The values of the two terms in the right-hand side of Eq.(\ref{approxd}) are given as
\begin{equation}
 \delta R_{\rm rel}\approx 0.190+0.091=0.280\,\mbox{fm},\qquad
 \delta R_{\rm non}\approx 0.101+0.074=0.175\,\mbox{fm}.\label{drr}
\end{equation}
Thus the difference about 0.1 fm between $\delta R$'s in the relativistic and non-relativistic models
mainly comes from the first terms $\epsilon R_p/3$,
and the diffuseness parameters yielding the second terms play
a rather minor role.
It should be noticed that the first term proportional to $\epsilon$ disappears,
when $\rho_{0,n}=(N/A)\rho$ and $ \rho_{0,p}=(Z/A)\rho$.

Since the values of $R_p$ and $\rho_{0,p}$ in the relativistic and non-relativistic
models are almost the same, the difference between $\delta R$'s in Eq.(\ref{approxd})
comes from $\rho_{0,n}$ in $\epsilon$.
Table 6 provides
\[
 \frac{(\rho_{0,n})_{\rm rel}}{(\rho_{0,n})_{\rm non}}=0.944, \qquad
  \frac{(\rho_{0,p})_{\rm rel}}{(\rho_{0,p})_{\rm non}}=0.995.
\]
The 5.6$\%$ decrease of $(\rho_{0,n})_{{\rm rel}}$ provides the increase of $(R_n)_{\rm rel}$
by $0.944^{-1/3}=1.019$, yielding the 0.1 fm-difference which we are discussing.
Fig.\ref{fig:NL3} shows qualitatively such a broadening of the neutron density in NL3, in comparing with that
in Fig.\ref{fig:SkMm}.

In Table 2 are listed the mean values of the neutron and proton densities for the
nuclear matter obtained by Eq.(\ref{me}).
Those values are almost the same as the corresponding ones in Table 6.
They provide the values of $\epsilon$ in Eq.(\ref{rn}) to be 0.1138 and 0.0567 for the relativistic
and non-relativistic models, respectively, which are comparable with the values for $^{208}$Pb in Table 6.
Thus, the various parameters including $\langle m^\ast_\tau\rangle$ and $\langle V_\tau\rangle$ at $r=0$
for $^{208}$Pb are similar to those for nuclear matter.

It may be noticed that the values listed in the rows of MF in Table 7 are a little different from those
of LSA in Table 4, since the former is the mean values of $R_\tau$ calculated by the models,
while the latter has been obtained by the least squares analysis of the calculated values
comparing with the experimental data\cite{kss}.

\section{The HVH lines in $^{208}$Pb}\label{hvhlines}

In the previous section, we have shown that the mean values of
$\langle m^\ast_\tau\rangle$, $\langle V_\tau\rangle$ and 
$\langle\rho_\tau\rangle$ in Table 2 for nuclear matter
are almost the same as the corresponding ones in Tables 5 and 6 for $^{208}$Pb.
In order to explore the reason why $\delta R$'s in the relativistic and non-relativistic
schemes are different from each other, the contributions from  $R_{{\rm ws},\tau}$ and
$a_{{\rm ws},\tau}$ in Figs.5 and 6 to $R_\tau$ should be also examined, in addition to those from
$V_{{\rm ws},\tau}$ and $m^\ast_{{\rm ws},\tau}$ in Figs.4 and 7.

In the present section, first, it will be discussed that the similar equations to  Eqs.(\ref{ra}) and (\ref{rra})
hold for the dependence of $R_\tau$ on the WS parameters and $m^\ast_{{\rm ws},\tau}$
with the values in Fig.4 to 7.
Second, the value of $\delta R$ will be shown to be dominated by $m^\ast_{{\rm ws},\tau}$
and $V_{{\rm ws},\tau}$, rather than by $R_{{\rm ws},\tau}$ and $a_{{\rm ws},\tau}$. 
Third, it will be investigated whether or not the constraint on the values of $m^\ast_{{\rm ws},\tau}$
and $V_{{\rm ws},\tau}$ by the HVH theorem holds in the same way as in Fig. 1 for the nuclear matter.
Finally, the difference between $\delta R$'s between the two schemes will be explained in terms of
$m^\ast_{{\rm ws},\tau}$ and $V_{{\rm ws},\tau}$.

We discuss $R_\tau$ of the finite nucleus $^{208}$Pb on the basis of Eqs.(\ref{ra}) and (\ref{rra}). For this purpose,
first we examine if it is appropriate for Eqs.(\ref{ra}) and (\ref{rra})
to employ $m^\ast_{{\rm ws},\tau}$ and $V_{{\rm ws},\tau}$ defined in Eqs.(\ref{nm}) and (\ref{wsp}).
When $R_\tau$ calculated by the simplified models is expressed in terms of
$m^\ast_{{\rm ws},\tau}$ and $V_{{\rm ws},\tau}$ as 
\begin{equation}
R_\tau\approx B_\tau\left(-\frac{R^2_{{\rm ws},\tau}}{m^\ast_{{\rm ws},\tau} V_{{\rm ws},\tau}}\right)^{1/4}
 (1+b_{{\rm ws},\tau})^{3/8}, \quad
 b_{{\rm ws},\tau} =\left(\frac{\pi a_{{\rm ws},\tau}}{R_{{\rm ws},\tau}}\right) ^2, \label{b}
\end{equation}
then the value of the coefficient $B_\tau$ corresponding to $B$ in Eq.(\ref{ra}) should be almost
constant independently of the various interaction parameters of the mean-field models.
In order to estimate the value of $B_\tau$,
both sides except for $B_\tau$ of the above equation are calculated for each model,
according to \S\ref{simp}.
The results are listed in the left-hand side named WS in Table 8,
where the mean values of $B_\tau$ are shown as $\langle B_\tau\rangle$ in units of
$({\rm fm}^2{\rm MeV})^{1/4}$ in the relativistic and non-relativistic models, separately.
The table shows that the values of the standard deviation($\sigma$) are small enough
for our purpose.

The meaning of $B_\tau$ may be qualitatively understood according to Ref.\cite{bm},
where the values of $C$ in Eq.(\ref{hr}) are estimated by summing a single particle radii
over the occupied orbits in HO potential.
Their approximations yield the values which are in the same
order of magnitude as those of $B_\tau$ in Table 8,
as $B_n\approx 5.44$ for $N=126$ and $B_p\approx 5.07$ for $Z=82$
in units of $({\rm fm}^2{\rm MeV})^{1/4}$.

\begin{table}
\begingroup
\renewcommand{\arraystretch}{1.2}
\hspace{4cm}%
{\setlength{\tabcolsep}{4pt}
\begin{tabular}{|c|c|c|c|} \hline
\multicolumn{2}{|c|}{} & 
\multicolumn{2}{c|}{$\langle B_\tau \rangle$ ($\sigma$)} \\ \cline{3-4}   
\multicolumn{2}{|c|}{}     & WS  & MF                 \\ \cline{1-4}
Rel    &\,n\, & 5.295 (0.0063) & 5.304 (0.0076) \\ \cline{2-4}
       &\,p\, & 5.243 (0.0151) & 5.253 (0.0151)\\ \cline{1-4}
Non    &\,n\, & 5.288 (0.0217) & 5.283 (0.0273) \\ \cline{2-4}
       &\,p\, & 5.268 (0.0801) & 5.261 (0.0685)\\ \hline
\end{tabular}
}
\endgroup
\caption{The mean values of the proportional constant $B_\tau$ between the root msr and the function of the
Woods Saxon parameters in Eq.(\ref{b}). The standard deviation $\sigma$ is given in the parenthesis.
WS lists the mean values in using $R_\tau$ of the simplified models, while MF shows for reference
those in using $R_\tau$ of the full mean-field approximation.
The numbers are written in units of $({\rm fm}^2{\rm MeV})^{1/4}$.
For details, see the text.}
\label{table_8}
\end{table}

We note that the values of $\sigma$ for protons are larger than those for neutrons
in both models.
This fact may be due to the Coulomb potential which is not explicitly taken into account
in the right-hand side in Eq.(\ref{b}).
If $v_c=22$ MeV is added to $V_{{\rm ws},p}$ by hand
for reference, the values of $\sigma$ for protons become comparable with those for neutrons,
as 0.0089 and 0.0217 in the relativistic and non-relativistic models, respectively.
We expect, however, that these results do not change the following discussions on the difference
between $\delta R$'s in the relativistic and non-relativistic models. 

It should  be also made sure that the 0.1 fm difference between $\delta R$'s is not
due to the enhancement of $R_{n,{\rm rel}}$ by the factor $B_{n,{\rm rel}}$. 
The equation corresponding to Eq.(\ref{rra}) is described as
\begin{equation}
\frac{R_n}{R_p} =\frac{B_n}{B_p}
 \left(\frac{m^\ast_{{\rm ws},p}V_{{\rm ws},p}}{m^\ast_{{\rm ws},n} V_{{\rm ws},n}}\right)^{1/4}
		\left(\frac{R_{{\rm ws},n}}{R_{{\rm ws},p}}\right)^{1/2}
		\left(\frac{1+b_{{\rm ws},n}}{1+b_{{\rm ws},p}}\right)^{3/8}\label{ratio}
\end{equation}
Since $\delta R$ is written as $\delta R=R_p(R_n/R_p-1)$
and the value of $R_p$ is almost fixed due to the fitting
in both relativistic and non-relativistic models,
the difference between $\delta R$'s
in the two frameworks stems from their values of $R_n/R_p$. Table 8 provides the ratio
of $(\langle B_n\rangle/\langle B_p\rangle)_{\rm rel}/(\langle B_n\rangle /\langle B_p\rangle)_{\rm non}
=1.0061$.
In the right-hand side of Table 8, the values of $\langle B_\tau\rangle$ in using $R_\tau$
from the mean-field calculations
in Eq.(\ref{b}) are listed. The table shows that the WS calculations yield almost the same results
as those of the mean-field ones, as
$(\langle B_n\rangle/\langle B_p\rangle)_{\rm rel}/(\langle B_n\rangle /\langle B_p\rangle)_{\rm non}=1.0055$.
These values imply that the factor $B_\tau$ is not enough to explain the 0.1 fm difference.
Thus, it is reasonable to use $V_{{\rm ws},\tau}$ and $m^\ast_{{\rm ws},\tau}$ defined
in Eqs.(\ref{wsp}) and (\ref{nm}) in the analysis of $\delta R$ in $^{208}$Pb.

Assuming that Eq.(\ref{ratio}) holds for the mean values in Table 5, we have
\begin{equation}
R_n \approx\frac{B_n}{B_p}
\left(\frac{\langle m^\ast_{{\rm ws},p}\rangle\langle V_{{\rm ws},p}\rangle}{\langle m^\ast_{{\rm ws},n}\rangle
V_{{\rm ws},n}\rangle}\right)^{1/4}
		\left(\frac{\langle R_{{\rm ws},n}\rangle}{\langle R_{{\rm ws},p}\rangle}\right)^{1/2}
		\left(\frac{1+\langle b_{{\rm ws},n}\rangle}{1+\langle b_{{\rm ws},p}\rangle}\right)^{3/8}R_p
		\label{mrms} \noindent
\end{equation}
with $\langle b_{{\rm ws},\tau}\rangle=(\pi\langle a_{{\rm ws},\tau}\rangle/\langle R_{{\rm ws},\tau}
\rangle)^2$. Using the values in Tables 5 and 8, the above equation provides
for the relativistic and non-relativistic models, respectively, as
\begin{equation}
R_{n,{\rm rel}}\approx 1.0517\,R_{p,{\rm rel}}\, , \quad
R_{n,{\rm non}}\approx 1.0274\,R_{p,{\rm non}}\, . \label{rmsr2} 
\end{equation}
If we put the numbers of $R_p$ obtained by the WS approximation in Table 7 
into the right-hand sides of Eq.(\ref{rmsr2}),
then we have the values of  $R_{n,{\rm rel}}\approx 5.739$ fm and $R_{n,{\rm non}}\approx 5.612$ fm.
It is seen that they are almost the same values as those of the WS approximation in Table 7. 
Thus, Eq.(\ref{ratio}) holds well also for the mean values in Tables 5 and 8.

Second, the value of $\delta R$ will be shown to be dominated by $m^\ast_{{\rm ws},\tau}$
and $V_{{\rm ws},\tau}$, rather than by $B_\tau$, $R_{{\rm ws},\tau}$ and $a_{{\rm ws},\tau}$,
with the use of their mean values. The one way to show this fact is by taking the numbers 
in Eq.(\ref{rmsr2}) which imply that $R_{n,{\rm rel}}>R_{n,{\rm non}}$,
when $R_{p,{\rm rel}}\approx R_{p,{\rm non}}$, indicating the $0.1$ fm difference problem.
Those numbers have been obtained by
\begin{equation}
1.0517\approx 1.0468 \times 1.0047\,, \quad
1.0274\approx 1.0329 \times 0.9947\,,\label{num} 
\end{equation}
where the first numbers in the right-hand sides of the above equations come from the factor
$(\langle m^\ast_{{\rm ws},p}\rangle\langle V_{{\rm ws},p}\rangle/\langle m^\ast_{{\rm ws},n}\rangle
V_{{\rm ws},n}\rangle)^{1/4}$, while the second numbers from the rest of the factors in Eq.(\ref{mrms}). 
Thus, the difference between $R_n$ and $R_p$ is mainly due to the first number coming from the values
of $\langle m^\ast_{{\rm ws},\tau}\rangle$ and $\langle V_{{\rm ws},\tau}\rangle$
in both relativistic and non-relativistic scheme.
The second numbers from $B_\tau$, $\langle R_{{\rm ws},\tau}\rangle$ and $\langle a_{{\rm ws},\tau}\rangle$
play a minor role in their differences.
This fact also implies that the distribution of $R_{{\rm ws},\tau}$
and $a_{{\rm ws},\tau}$ over a wide region in Figs.5 and 6 is not worrisome for the 0.1 fm problem.
The minor role of $a_{{\rm ws},\tau}$ is consistent with the results of Eq.(\ref{drr}).

It may be seen in another way qualitatively that,
compared to $B_\tau$, $\langle R_{{\rm ws},\tau}\rangle$
and $\langle b_{{\rm ws},\tau}\rangle$, $\langle m^\ast_{{\rm ws},\tau}\rangle$
and $\langle V_{{\rm ws},\tau}\rangle$ play an important roles in $\delta R$
of the two frameworks.
We write Eq.(\ref{b}) in terms of the mean values,
\begin{equation}
R_{\tau,{\rm rel}}\approx B_{\tau,{\rm rel}}\left(-(\langle m^\ast_{{\rm ws},\tau}\rangle
 \langle V_{{\rm ws},\tau}\rangle)_{\rm rel}\right)^{-1/4}
 (\langle R_{{\rm ws},\tau}\rangle_{\rm rel})^{1/2}
 (1+\langle b_{{\rm ws},\tau}\rangle_{\rm rel})^{3/8},  \label{mb}
\end{equation}
for the relativistic scheme. In the above equation, keeping the values
of $B_{\tau,{\rm rel}}$, $\langle R_{{\rm ws},\tau}\rangle_{\rm rel}$
and $\langle b_{{\rm ws},\tau}\rangle_{\rm rel}$,
we replace $(\langle m^\ast_{{\rm ws},\tau}\rangle \langle V_{{\rm ws},\tau}\rangle)_{\rm rel}$
by that of the non-relativistic one,
$(\langle m^\ast_{{\rm ws},\tau}\rangle \langle V_{{\rm ws},\tau}\rangle)_{\rm non}$.
Then, the values of $R_{\tau,{\rm rel}}$ of the WS approximation in Table 7
and the mean values of Table 5 provide
\begin{equation}
R_{\tau,{\rm rel}}\left(\frac{\bigl(\langle m^\ast_{{\rm ws},\tau}\rangle
\langle V_{{\rm ws},\tau}\rangle\bigr)_{\rm rel}}
		{\bigl(\langle m^\ast_{{\rm ws},\tau}\rangle \langle V_{{\rm ws},
\tau}\rangle\bigr)_{\rm non}}\right)^{1/4}
=
\left\{
\begin{array}{ll}
5.607\,\mbox{fm}, & \tau=n, \\[2pt]
5.403\,\mbox{fm}, & \tau=p. 
\end{array}
\right.
\label{mv}
\end{equation}
%Thus, the replacement of $(\langle m^\ast_{{\rm ws},\tau}\rangle
%\langle V_{{\rm ws},\tau}\rangle)_{\rm rel}$
%by $(\langle m^\ast_{{\rm ws},\tau}\rangle \langle V_{{\rm ws},\tau}\rangle)_{\rm non}$
%in $R_{\tau,{\rm rel}}$
%makes the values of $R_{\tau,{\rm rel}}$ smaller than those of $R_\tau$
%of the non-relativistic scheme in Table 7.
The above equation yield $\delta R=5.607-5.403=0.204 $\,fm,
which should be compared to $\delta R =0.159$\,fm in WS for the non-relativistic models in Table 7.
The difference between $\delta R$'s in the two frameworks is reduced
from $0.283-0.159=0.124$\,fm to $0.204-0.159=0.045$\,fm by $64\%$.
Thus, it is seen that the $0.1$ fm problem is deeply related to
the difference between $(\langle m^\ast_{{\rm ws},\tau}\rangle
\langle V_{{\rm ws},\tau}\rangle)_{\rm rel}$
and $(\langle m^\ast_{{\rm ws},\tau}\rangle \langle V_{{\rm ws},\tau}\rangle)_{\rm non}$ in $R_\tau$ .

\begin{figure}[ht]
\begin{center}
\includegraphics[scale=1]{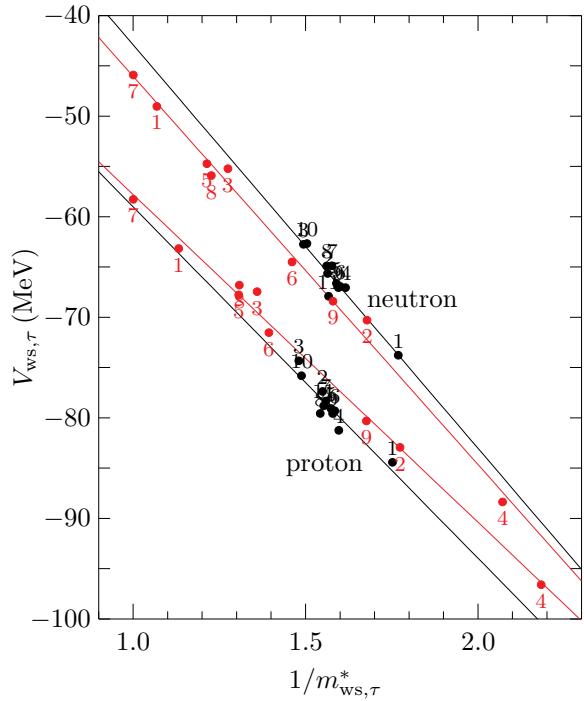}%
\caption{The relationship between the effective mass and the one-body potential
for neutrons($\tau=n$) and protons ($\tau=p$) in the mean-field models for $^{208}$Pb.
The black circles show the values calculated by the 11 relativistic models,
while the red ones those by the 9 non-relativistic models.
The least-square lines are shown
for the four groups. The two black lines are obtained from the black circles
for neutrons and protons, respectively, and the red lines from the red circles. 
}
\label{fig:mws_Vws}
\end{center}
\end{figure}

Third, let us investigate whether or not there is a constraint on $m^\ast_{{\rm ws},\tau}$
and  $V_{{\rm ws},\tau}$ in $^{208}$Pb, as in nuclear matter.
In Fig.\ref{fig:mws_Vws} are plotted the values of $V_{{\rm ws},\tau}$ in Fig.\ref{fig:Vws_force}
and those of $m^\ast_{{\rm ws},\tau}$ in Fig.\ref{fig:mws_force}
in the $1/m^\ast_{{\rm ws},\tau}-V_{{\rm ws},\tau}$ plane.
The black circles show the values for neutrons and protons in the relativistic models,
while the red circles in the non-relativistic models.
The numbers attached to each circle indicate the used model, according to
the numbering mentioned in \S \ref{asnm}.
The pair of the same number represent the values for neutrons and protons calculated by the same model.

The slanting lines are obtained by the least square method for the values of each group.
The upper and lower black lines are drawn for neutrons and protons in the
relativistic models, respectively, while the upper and lower red lines are in the
non-relativistic models.
It is remarkable that the values of each group follow well the corresponding line, and that
the four lines are well separated from one another, as in Fig.\ref{fig:asym_v_em} for nuclear matter.
We notice that  the only FSU(11)\cite{fsu} among the relativistic models yields
a point on the neutron line for the non-relativistic models. 
This may reflect the fact that FSU has added two additional parameters to the Lagrangian of,
for example, NL3(5)\cite{nl3}, so as to reduce the value of $R_n$.
The values of the gradient($a^{\rm L}_{{\rm ws},\tau}$) and the intercept($b^{\rm L}_{{\rm ws},\tau}$)
of the LSL,
\begin{equation}
V_{{\rm ws},\tau}=a^{\rm L}_{{\rm ws},\tau}/m^\ast_{{\rm ws},\tau}+b^{\rm L}_{{\rm ws},\tau},\label{wshvh}
\end{equation}
are listed in Table 9 for relativistic(Rel) and non-relativistic(Non) models. 
The values of the correlation coefficient, $r$, are also shown, which are nearly equal to 1.

\begin{table}
\begingroup
\renewcommand{\arraystretch}{1.2}
\hspace{3.6cm}%
{\setlength{\tabcolsep}{4pt}
\begin{tabular}{|c|c|c|c|c|c|c|} \hline
     &
\multicolumn{3}{c|}{Rel} &
\multicolumn{3}{c|}{Non} \\ \cline{2-7}\hline
    &
$a^{\rm L}_{{\rm ws},\tau}$  &
$b^{\rm L}_{{\rm ws},\tau}$  &
$r$  & 
$a^{\rm L}_{{\rm ws},\tau}$  &
$b^{\rm L}_{{\rm ws},\tau}$  &
$r$   \\ \cline{2-7}\hline
$n$   & $-40.120$ & \hphantom{2}$-2.795$ & $0.958$& $-38.609$&\hphantom{2}$-7.438$ & $0.997$  \\ \hline
$p$   & $-34.933$ & $-24.098$ & $0.937$ & $-32.588$ &$-25.215$ & $0.997$  \\ \hline 
\end{tabular}
}
\endgroup
\caption{The values of the gradient($a^{\rm L}_{{\rm ws},\tau}$) and
 the intercept($b^{\rm L}_{{\rm ws},\tau}$) of the least-square line
for the relationship between the depth of the one-body potential and the nucleon effective mass in Fig.\ref{fig:mws_Vws}.
The numbers of $a^{\rm L}_{{\rm ws},\tau}$ and $b^{\rm L}_{{\rm ws},\tau}$ are given in units of MeV.
The values of the correlation coefficient, $r$, are also listed.
For details, see the text.}
\label{table_9}
\end{table}

In Fig.\ref{fig:mws_Vws}, it is seen that the variation of the effective mass and the strength
of the one-body potential
in finite nuclei is also constrained  in a similar way to that in Fig.\ref{fig:asym_v_em} for nuclear matter.
Eq.(\ref{wshvh}) has the same form as in Eq.(\ref{hvhline}) from the HVH theorem.
Thus, the HVH theorem seems to be inherent in the mean field approximation for finite nuclei also.
From now on, we will refer to the LSL of $^{208}$ Pb as the HVH line. 

We compare the coefficients of the HVH line 
for $^{208}$Pb in Table 9 to those of Eq.(\ref{hvhline})
for nuclear matter listed in Table 2.  
The coefficients of Eq.(\ref{hvhline}) have shown to be constrained by the HVH theorem through
Eq.(\ref{hvhav}).
It is seen in Tables 2 and 9 that the values of the corresponding coefficients are not the same
as each other, but the magnitude relation of the corresponding two values are almost the same as
those of other pair.
More important values for the present discussion is those in Eq.(\ref{hvhav}).
In Table 10 is listed one of the values in Eq.(\ref{hvhav}),
$\langle a_{\tau} \rangle+\langle b_{\tau}\rangle\langle m^\ast_{\tau}\rangle$,
in the first column, 
and the corresponding values obtained from Tables 5 and 9 in the second and the last column. 
It is seen that the values for nuclear matter in the first column are almost the same as those for $^{208}$Pb
in other columns.
Since the values in the first column is nothing but the results due to the HVH theorem, it is 
confirmed that those in the second and third columns also reflect the constraint by the theorem.

We note that the values of the first column has been obtained by introducing
a model with $v_c$
in Eq.(\ref{med}). It is made so as to provide neutrons and protons with the same average
binding energies by Eq.(\ref{me}), as in stable nuclei. 
The value of $v_c $ is employed which approximately corresponds to the energy of the
Coulomb potential for $^{208}$Pb in Eq.(\ref{coulomb}). 
Although the model has been used as a guide for discussions of the finite nucleus,
Table 10 shows conversely that such a simple model almost reproduces results for the finite nucleus
and is useful for describing asymmetric nuclear matter.

In Fig.\ref{fig:mws_Vws}, it should be noticed that 
on the one hand, that the distance between two black lines for the relativistic models
is about 12.5 MeV at a fixed value of $1/m^\ast_{{\rm ws},\tau}$, as listed in Table 11.
It is almost the same as the mean value of $V_3=V_{{\rm ws},n}-V_{{\rm ws},p}$
in Fig.\ref{fig:Vws_force}. This is
because of $m^\ast_{{\rm ws},n}\approx m^\ast_{{\rm ws},p}$ in the relativistic models. 
On the other hand, in the non-relativistic models, 
the distance between the two red lines for the same value of $1/m^\ast_{{\rm ws},\tau}$
is about 7.7 MeV, in spite of the fact that $V_3\approx 11.6$ MeV as in Table 11.
This is because, in the non-relativistic models, the value of the neutron effective mass is larger than
that of the proton one, except for SLy4(6)\cite{sly4}, as in Fig.\ref{fig:mws_force}.
The value of $V_3\approx 11.6$ MeV is approximately kept by providing neutrons and protons
with different effective masses. As seen below, it is essential for understanding 0.1 fm difference
that the values of the effective mass for neutrons are different from the ones for protons
in the  non-relativistic models, while those in the relativistic models are almost the same.

So far, understanding the dependence of $R_\tau$ on
$m^\ast_{{\rm ws},\tau}$ and $V_{{\rm ws},\tau}$ is simply based on their mean values as in Eqs.(\ref{num})
and (\ref{mv}), aiming to emphasize their roles in the $0.1$ fm problem.
 Finally, we investigate roles of of $m^\ast_{{\rm ws},\tau}$ and $V_{{\rm ws},\tau}$ in $R_\tau$
 by using their values themselves, together with the HVH line in Fig.{\ref{fig:mws_Vws}.

Eq.(\ref{mv}) has been obtained by replacing the mean values,
$\left(\langle m^\ast_{{\rm ws},\tau}\rangle \langle V_{{\rm ws},\tau}\rangle\right)_{\rm rel}$
by $\left(\langle m^\ast_{{\rm ws},\tau}\rangle \langle V_{{\rm ws},\tau}\rangle\right)_{\rm non}$.
In order to explore in more detail how the value of $R_\tau$ in the relativistic scheme approaches
to that in the non-relativistic one by changing $m^\ast_{{\rm ws},\tau}$ and $V_{{\rm ws},\tau}$, 
%we make a similar replacement in each model.
%according to the HVH line as follows.
we replace the values of  $m^\ast_{{\rm ws},\tau}$ and $V_{{\rm ws},\tau}$ in $R_\tau$ in
each relativistic model by $\langle m^\ast_{{\rm ws},\tau}\rangle_{\rm non}$
and $\langle V_{{\rm ws},\tau}\rangle_{\rm non}$.
The replacement will be made keeping the values
of $R_{{\rm ws},\tau}$ and $b_{{\rm ws},\tau}$ in each model, and using
the HVH line in Fig.\ref{fig:mws_Vws}. By this procedure, we will see the roles
of $m^\ast_{{\rm ws},\tau}$ and $V_{{\rm ws},\tau}$ in $R_\tau$ separately, as follows.

In Fig.\ref{fig:NL3-Rtau-mws} is shown $R_\tau$ as a function of
$m^\ast_{{\rm ws},\tau}$ in the case of NL3(5) as an example.
The closed and open circles indicate the values of
$R_\tau$ in the full mean-field calculation and in the simplified one in \S\ref{simp}, respectively,
at the value of $m^\ast_{{\rm ws},\tau}$ for NL3.
The solid curves are calculated by keeping the values of $R_{{\rm ws},\tau}$ and $a_{{\rm ws},\tau}$
of NL3 and using $V_{{\rm ws},\tau}$ given by the HVH line
for the relativistic models in Fig.\ref{fig:mws_Vws}.
The closed and open circles are seen to be almost on the curves.
The dashed curves also show $R_\tau$, but using  $V_{{\rm ws},\tau}$ given by the HVH line
for the non-relativistic models in Fig.\ref{fig:mws_Vws}.

In Fig.\ref{fig:NL3-Rtau-mws}, we have specified the six points on the curves,
where X$_n$, Y$_n$ and Z$_n$ are for the neutrons, and others for the protons.
The points X$_\tau$ indicates the position of the open circles.
The points Y$_\tau$ and Z$_\tau$ are on the dashed curves. The former
indicates the place where NL3 provides $m^\ast_{{\rm ws},\tau}$, and the latter 
the place of $m^\ast_{{\rm ws},\tau}=\langle m^\ast_{{\rm ws},\tau}\rangle_{\rm non}$
as shown by the vertical lines.
The replacement of $m^\ast_{{\rm ws},\tau}$ and $V_{{\rm ws},\tau}$ in NL3
by $\langle m^\ast_{{\rm ws},\tau}\rangle_{\rm non}$ and $\langle V_{{\rm ws},\tau}\rangle_{\rm non}$
is made by using the values at point $Z_\tau$.
%instead of those at $X_n$.
We make, however, the replacement by two steps according to the curves in Fig.\ref{fig:NL3-Rtau-mws}.
%We explore a change of $R_\tau$ due to $m^\ast_{{\rm ws},\tau}$ and $V_{{\rm ws},\tau}$,
%as in  Eq.(\ref{mv}), but by the two steps, according to the curves in Fig.9\ref{fig:NL3-Rtau-mws}.
In the first step, the values at X$_\tau$ are replaced by those at Y$_\tau$, and in the second step
the values at Y$_\tau$ by those at Z$_\tau$.
This process is shown in Fig.\ref{fig:NL3-Rtau-mws} by the arrows.
The blue arrow indicates the first step, while the green one the second step.
In this way, we may see how $R_\tau$ in NL3 varies
by $m^\ast_{{\rm ws},\tau}$ and $V_{{\rm ws},\tau}$ separately and approaches to $R_\tau$ in
the non-relativistic scheme.

Fig.\ref{fig:NL3-Rtau-mws} shows that the value of $R_n$ is decreased in the first step,
because $V_{{\rm ws},n}$ becomes deeper as seen in Fig.\ref{fig:mws_Vws}. 
From Y$_n$ to Z$_n$, the potential becomes shallower, 
but the value of the effective mass is increased and a role of the kinetic part as a repulsive
potential declines.
As a result, the value of $R_n$ further shrinks, as in Fig.\ref{fig:NL3-Rtau-mws}.
The decrease of $R_\tau$ from Y$_{\tau}$ to Z$_{\tau}$ with the increasing $m^\ast_{{\rm ws},\tau}$
is understood qualitatively by Eqs.(\ref{b}) and(\ref{wshvh}) which yield
\begin{equation}
m^\ast_{{\rm ws},\tau}
 =-\,\frac{1}{b^{\rm L}_{{\rm ws},\tau}}
 \left(a^{\rm L}_{{\rm ws},\tau}
+\frac{B_\tau^4}{R_\tau^4} R^2_{{\rm ws},\tau}\bigl(1+b_{{\rm ws},\tau}\bigr)^{3/2}\right).\label{mrv}
\end{equation}
Thus, the value of $R_n$ in the relativistic models approaches to that in the non-relativistic
models, following the path under the constraint of the HVH theorem on  $V_{{\rm ws},\tau}$
and $m^\ast_{{\rm ws},\tau}$.

With respect to $R_p$, Fig.\ref{fig:NL3-Rtau-mws} shows its increase from X$_p$ to Y$_p$,
because of the decreasing strength of $|V_{{\rm ws},p}|$.
From Y$_p$ to Z$_p$, the value of $R_p$ decreases in the same way
as that of $R_n$ from Y$_n$ to Z$_n$, according to Eq.(\ref{mrv}).
The final value of $R_p$ at the point Z$_p$ returns
to the almost original value at X$_p$,
since the value of $R_p$ at X$_p$ for the relativistic model is fixed by the experimental value
of $R_c$ as an input,
while the value at Z$_p$ is almost equal to the values of $R_p$ for the non-relativistic models
which are fixed in the same way.

In the above analysis, it should be noticed
that the value of $\langle m^\ast_{{\rm ws},n}\rangle_{\rm non}$
is larger than that of $\langle m^\ast_{{\rm ws},p}\rangle_{\rm non}$ as in Table 5. 
Owing to the fact, the path from Y$_n$ to Z$_n$ is longer than that from Y$_p$ to Z$_p$
as seen in Fig.\ref{fig:NL3-Rtau-mws}. 
This difference also works to make $R_{n}$ smaller in the path from Y$_n$ to Z$_n$.

Fig.\ref{fig:Rtau-force} shows the values of $R_\tau$ which are obtained
by the same procedure as in Fig.\ref{fig:NL3-Rtau-mws}
for all the relativistic mean-field models taken in the present paper.
The black and red circles show the results of the full mean-field calculations and the simplified
ones in \S\ref{simp}, respectively, where the closed circles are for neutrons and the open circles
for protons. The vertical lines indicate their mean values. It is seen that the simplified
calculations reproduce well the values of $R_\tau$ by the full calculations.
Those for the non-relativistic models also are shown in the same way.

\begin{table}
\begingroup
\renewcommand{\arraystretch}{1.2}
\hspace{2.6cm}%
{\setlength{\tabcolsep}{4pt}
\begin{tabular}{|c|c|c|c|c|} \hline
\multicolumn{2}{|c|}{}     &
$\langle a_{\tau} \rangle + \langle b_{\tau}\rangle\langle m^\ast_{\tau}\rangle$  &
$\langle m^\ast_{{\rm ws},\tau}\rangle\langle V_{{\rm ws},\tau}\rangle$ &
$a^{\rm L}_{{\rm ws},\tau} + b^{\rm L}_{{\rm ws},\tau}\langle m^\ast_{{\rm ws},\tau}\rangle$  \\ \hline   
${\rm Rel}$   & $n$  & $-41.814$ & $-41.956$ & $-41.887$ \\ \cline{2-5}
              & $p$  & $-49.737$ & $-50.386$ & $-50.324$ \\ \hline
${\rm Non}$   & $n$  & $-44.321$ & $-46.075$ & $-44.193$ \\ \cline{2-5}
              & $p$  & $-51.364$ & $-52.437$ & $-50.760$  \\ \hline 
\end{tabular}
}
\endgroup
\caption{The values for the combination composed of the effective mass and the coefficients of the least-square line
 for asymmetric nuclear matter(the first column) and for $^{208}$Pb(the third column)
 in the relativistic and non-relativistic models.
The values of the product of the effective mass and the strength of the one-body potential is also listed
in the second column for $^{208}$Pb. All the numbers are given in units of MeV.
For details, see the text.}
\label{table_10}
\end{table}

\begin{table}
\begingroup
\renewcommand{\arraystretch}{1.2}
\hspace{3cm}%
{\setlength{\tabcolsep}{4pt}
\begin{tabular}{|c|c|c|c|} \hline
     &
$V^{\rm rel}_{{\rm ws},\tau}\bigl(\langle m^\ast_\tau\rangle_{\rm rel}\bigr)$ &
$V^{\rm non}_{{\rm ws},\tau}\bigl(\langle m^\ast_\tau\rangle_{\rm rel}\bigr)$   &
$V^{\rm non}_{{\rm ws},\tau}\bigl(\langle m^\ast_\tau\rangle_{\rm non}\bigr)$ \\ \hline 
$n$   & $-66.264$  & $-68.517$ & $-58.865$  \\ \cline{1-4}
$p$   & $-78.792$  & $-76.238$ & $-70.434$  \\ \hline
$V_n-V_p$ & \hphantom{$-$}$12.528$  & \hphantom{$-7$}$7.721$ & \hphantom{$-$}$11.569$  \\ \cline{1-4}   \hline 
\end{tabular}
}
\endgroup
\caption{The neutron(n) and proton(p) potentials at the mean value of the effective mass.
The numbers are given in units of MeV.
 For details, see the text.}
\label{table_11}
\end{table}

\begin{figure}[ht]
\begin{center}
\includegraphics[scale=1]{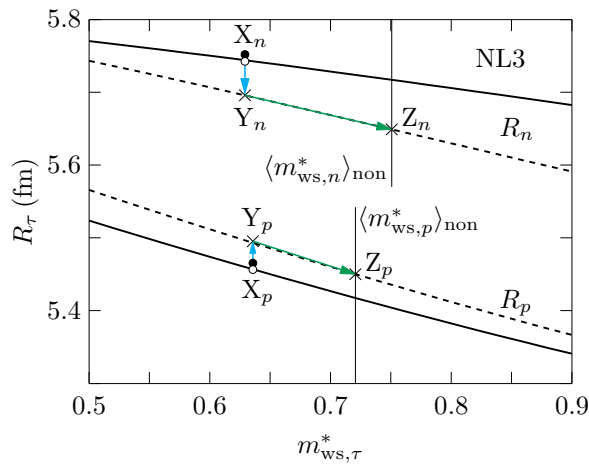}%
\caption{The root msr as a function of the effective mass for $\tau=n$ and $\tau=p$ in NL3.
The closed circles are obtained by the full mean-field approach, and the open ones by the simplified
Hamiltonian of NL3. The solid curves are calculated by using the least-square lines
for the relationship between $m^\ast_{{\rm ws},\tau}$ and $V_{{\rm ws},\tau}$
in the relativistic models in Fig.\ref{fig:mws_Vws},
while the dashed ones by using those in the non-relativistic models.
The top two lines are for neutrons, and the bottom two ones for the protons.
In these curves, the values of the Woods-Saxon parameters, $R_{{\rm ws},\tau}$
and $a_{{\rm ws},\tau}$, are taken from those determined by NL3.
The vertical lines indicate the average values of the effective masses 
for neutrons($\langle m^\ast_{{\rm ws},n}\rangle_{\rm non}$)
and protons($\langle m^\ast_{{\rm ws},p}\rangle_{\rm non}$) in the non-relativistic models.
The point, X$_n$, indicates the place of the open circle for the neutrons,
while Y$_n$ the point on the dashed curve at the same value of $m^\ast_{{\rm ws},n}$ as that for X$_n$.
The point, Z$_n$ shows the intersection point between the dashed curve and the vertical line
for $\langle m^\ast_{{\rm ws},n}\rangle_{\rm non}$.  
The points, X$_p$, Y$_p$ and Z$_p$ are given in a similar way.
The blue and green arrows are used for discussions in the text.
}
\label{fig:NL3-Rtau-mws}
\end{center}
\end{figure}

\begin{figure}[ht]
\begin{minipage}[t]{7.2cm}
%\begin{center}
\includegraphics[scale=0.9]{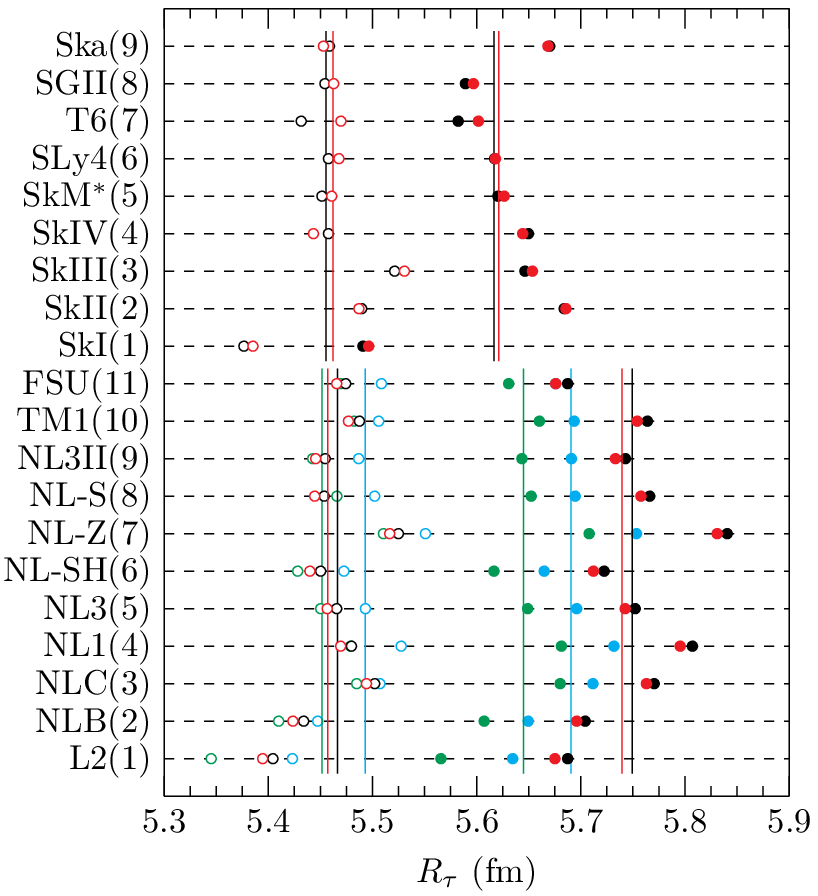}%
\caption{The root msr $R_\tau$ of $^{208}$Pb calculated in the relativistic and non-relativistic models.
 On the left-hand side, the used models are indicated. The black closed and open circles are obtained
 by the full mean-field approximations for the neutrons($\tau=n$) and protons($\tau=p$), respectively.
 The red ones are by the simplified Hamiltonian for each model. The blue ones are  obtained
 through the first step discussed in the text, while the green ones by the second step.
 The vertical lines show the average values of the same color circles, respectively.
For details, see the text.}
\label{fig:Rtau-force}
%\end{center}
%\end{figure}
\end{minipage}\hspace{0.8cm}%
\begin{minipage}[t]{7.2cm}
%\begin{figure}[ht]
%\begin{center}
\includegraphics[scale=0.9]{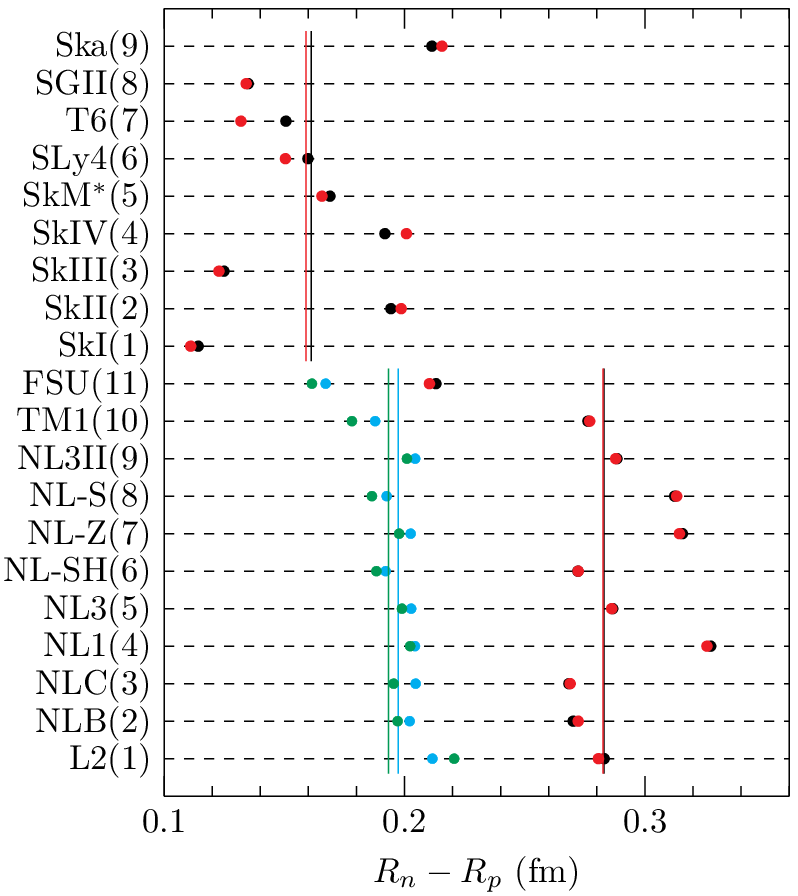}%
\caption{The difference between the root msr's of the neutron and the proton
 distribution in $^{208}$Pb calculated in the relativistic and non-relativistic models.
On the left-hand side, the used models are indicated. The black closed circles are obtained
 by the full mean-field approximations, and the red ones by the simplified Hamiltonian for each model.
 The blue ones are obtained through the first step, while the green ones through the second step
 discussed in the text.  The vertical lines show the average values of the same color circles,
 respectively. For details, see the text.}
\label{fig:RnRp-force}
%\end{center}
\end{minipage}
\end{figure}

The blue circles are obtained by the first step from X$_\tau$ to Y$_\tau$
mentioned in Fig.\ref{fig:NL3-Rtau-mws},
while the green ones by the second step. All the models show the similar change of $R_\tau$
as in Fig.\ref{fig:NL3-Rtau-mws} such that the values of $R_n$  decrease by the two steps,
while those of $R_p$
come back to almost the same values by the second step from Y$_\tau$ to Z$_\tau$.

Fig.\ref{fig:RnRp-force} shows the values of $\delta R$, using the same designating symbols as in Fig.\ref{fig:Rtau-force}.
The values from the two steps shown by the green circles are almost the same as the blue ones
obtained by the first step, since the values of $R_p$ return to the original ones by the
second step.

The results of $R_\tau$ and $\delta R$ in Figs.\ref{fig:Rtau-force} and \ref{fig:RnRp-force} are
summarized in Table 12 in units of fm.
The mean values of $R_\tau$ in the relativistic models are listed in the columns named Red, Blue
and Green according to the colors of those figures. From Red to Green, the value of $R_n$ decreases,
while that of $R_p$ increases from Red to Blue and decreases from Blue to Green, up to
almost the Red one, as shown in the figures. In changing the values $V_\tau$ and $m^\ast_\tau$
in the relativistic models following the HVH lines, the value of $\delta R$ shrinks 
from 0.283 fm to 0.193 fm, which should be compared to 0.159 fm of the non-relativistic models. 
The difference between $\delta R$'s in the relativistic and non-relativistic models becomes smaller
by $73\%$, changing its value from 0.124 fm to 0.034 fm.

\begin{table}
\begingroup
\renewcommand{\arraystretch}{1.2}
\hspace{4.5cm}%
{\setlength{\tabcolsep}{4pt}
\begin{tabular}{|c|c|c|c|c|} \hline
 & \multicolumn{3}{c|}{Rel} & Non \\ \hline
  & Red  & Blue  & Green  & Red  \\ \hline
 $R_n$      &$5.740$ &5.691 & 5.645 &$5.621$ \\ \cline{1-5}
 $R_p$      &5.457   &5.493 & 5.452 &5.462  \\ \hline
 $\delta R$ &$0.283$ &0.197 & 0.193 &0.159 \\ \cline{1-5}   \hline 
\end{tabular}
}
\endgroup
\caption{
 The average values of the root msr of the proton and neutron distributions in the relativistic(Rel)
 and non-relativistic(Non) models in various approximations. Red, Green and Blue indicate the
 average values corresponding to those in Figs.\ref{fig:Rtau-force} and \ref{fig:RnRp-force}, respectively.
All the numbers are given in units of fm. For details, see the text.}
\end{table}

It is concluded that most of the 0.1 fm difference between $\delta R$'s
in the relativistic and non-relativistic models is attributed to the difference between the values of 
their $V_\tau$ and $m^\ast_\tau$, which are constrained by $\rho_{0,\tau}$ through the HVH theorem.
The remaining difference may be caused by the sum of many small contributions, in addition to
those from $R_{{\rm ws},\tau}$, $a_{{\rm ws},\tau}$ and $B_\tau$, from the used approximations.
The exchange term of the Coulomb force, the center of mass
correction, the small component of the wave functions, etc., are also managed differently in the two frameworks.
Discussions on those effects, however, are beyond the present purpose.

\section{Summary}

Ref.\cite{kss} has pointed out that the neutron skin thickness defined
by $\delta R=R_n-R_p$ in $^{208}$Pb is larger by about 0.1 fm in the relativistic mean-field models
than in the non-relativistic ones. 
Here, $R_n$ and $R_p$ represent the root msr(mean-square-radius)
of the point neutron and proton distributions
in the nucleus, respectively.
The value of the charge radius $R_c$ of $^{208}$Pb is about 5.503 fm.
The 0.1 fm difference is not small for nuclear physics\cite{bm,suda,sly,hargen},
but also for astrophysics\cite{thi,adh,hargen}.
In this paper, it has been investigated why the difference is avoidable
in the present mean-field models, even though both relativistic ad non-relativistic
models are constructed phenomenologically with free parameters to be fixed by experimental values.

The value of $R_p$ is one of the most important inputs,
together with the binding energy per nucleon and the Fermi momentum in nuclear matter
in all of the phenomenological models\cite{sw, sly4}.
The relationship between $R_p$ and $R_c$ is unambiguously defined theoretically\cite{ksmsr},
and the latter is observed experimentally through electromagnetic probes, whose reaction mechanism are
well understood\cite{deforest, vries,bd}.
Hence, the 0.1 fm problem is due to the difference between the values of $R_n$ in the two frameworks.

It is shown that the values of $R_\tau$ are dominated by those of $(-m^\ast_{\tau}V_\tau)^{-1/4}$,
as in Eq.(\ref{b}),
where $m^\ast_\tau$ and $V_\tau$ represent the effective mass in units of $M$ and
the strength of the one-body potential near the center of the nucleus($r\approx 0$),
respectively, and
the subscript indicates $\tau=p$ for protons and $\tau=n$ for neutrons.
Although $m^\ast_\tau$ and $V_\tau$ are complicated functions of the interaction parameters
in the phenomenological models, they are not independent of each other. Their variations
are constrained together with the nucleon density $\rho_\tau$ at $r\approx 0$
by the Hugenholtz-Van Hove(HVH)
theorem\cite{bethe, weisskopf, hvh}.

In writing the average values of $m^\ast_{\tau}$ and $V_\tau$ in each framework as
 $\langle m^\ast_{\tau}\rangle$ and $\langle V_\tau\rangle$, respectively,
their product is approximately expressed by the HVH equation as
$\langle m^\ast_{\tau}\rangle\langle V_\tau\rangle
\approx a_\tau+b_\tau\langle m^\ast_{\tau}\rangle$, where $a_\tau$ and $b_\tau$ are constants.
The values of $a_\tau$ and $b_\tau$ depend on the average values of
$\rho_\tau$($\langle \rho_{\tau}\rangle$),
the binding energy per nucleon $\eb$ and Coulomb energy $v_c$
of the corresponding asymmetric nuclear matter with $N$ and $Z$.
Since the values of $\eb$ and $v_c$ are almost the same in the relativistic
and non-relativistic models, the difference between the two frameworks in the right-hand side
of the HVH equation is attributed to the difference between the values of $\langle \rho_\tau\rangle$
and $\langle m^\ast_\tau\rangle$.
Indeed, the values of $\langle\rho_\tau\rangle$ and $\langle m^\ast_\tau\rangle$ in the nuclear matter
in Table 2 are almost the same as those for $^{208}$Pb in Table 5 and 6.
The difference of the right-hand side of the HVH equation for the two frameworks
is expressed by $\langle m^\ast_{\tau}\rangle\langle V_\tau\rangle$ in the left-hand side,
which induces the difference of $R_n$ between the relativistic and non-relativistic models,
according to Eq.(\ref{b}).

Table 10 provides their average values as  $(\langle m^\ast_n\rangle\langle V_n\rangle)_{\rm non}
 =-46.075$\,MeV for the non-relativistic models
against $(\langle m^\ast_n \rangle \langle V_n \rangle)_{\rm rel}=-41.956$\,MeV for the relativistic  models.
The ratio of these values yields 
\[
(46.075/41.956)^{1/4}=1.0237,
\]
which is comparable to the value
showing the 0.1 fm difference of $R_n$ as
\[
R_{n,{\rm rel}}/R_{n,{\rm non}}=5.740/5.621=1.0212
\]
in Table7.
This comparison assumes the same relationship between the average values of
$R_n$ and  $(-m^\ast_nV_n)^{-1/4}$, as in Eq.(\ref{b}).
The results of the more detailed analysis without using the average values have been summarized
in Table 12, which shows that about $70\%$ of the 0.1 fm difference is explained according to the HVH theorem.

We note that the 0.1 fm problem is observed using the limited number of
the Skyrme-type interactions and the relativistic mean-field models in Ref.[4], so that
the problem has been investigated within the same models in the present paper.
It may be interesting to explore in other phenomenological models[2]
whether or not there is a similar difference problem
and the HVH theorem is useful for understanding  the difference.

 The 0.1 fm difference has been observed in $^{48}$Ca also in Ref.\cite{kss}.
It would be discussed in a similar way as for $^{208}$Pb in the present paper, 
but a new method must be devised for comparing the results for $^{48}$Ca 
with those for nuclear matter in the mean-field models.

\section*{Acknowledgment}
The authors would like to thank Professor T. Suda for useful discussions.

\end{document}